\documentclass[
prd,
superscriptaddress,
longbibliography,
nofootinbib,
amsmath,
amssymb
]{revtex4-2}

\usepackage{mathtools}
\usepackage{bm}
\usepackage[normalem]{ulem}

\usepackage{graphicx}
\usepackage{xcolor}
\usepackage{booktabs}
\usepackage{array}

\usepackage{tikz}
\usetikzlibrary{arrows.meta,calc,positioning}

\usepackage[
colorlinks=true,
linkcolor=blue,
citecolor=blue,
urlcolor=blue
]{hyperref}

\usepackage[nameinlink,noabbrev]{cleveref}

\newcommand{\kpar}{k_{\parallel}}
\newcommand{\kq}{\bm{k}_q}
\newcommand{\aN}{\alpha_N}
\newcommand{\vF}{v_F}
\newcommand{\half}{\tfrac12}
\newcommand{\Rdisk}{\mathcal{R}}
\newcommand{\rA}{\bm r_A}
\newcommand{\Pt}{P^{t}}
\newcommand{\Pl}{P^{l}}

\DeclareMathOperator{\Imag}{Im}

\begin{document}

\title{Lamb Shift of a Static Atom Facing a Rotating Surface}

\author{C\'esar D. Fosco}
\affiliation{Instituto Balseiro, Centro Atómico Bariloche, San Carlos de Bariloche, Argentina}

\author{Fernando C. Lombardo}
\affiliation{Departamento de Física ``Juan José Giambiagi'', Facultad de Ciencias Exactas y Naturales, Universidad de Buenos Aires, Buenos Aires, Argentina and Instituto de Física de Buenos Aires (IFIBA), CONICET--Universidad de Buenos Aires, Buenos Aires, Argentina}

\author{Francisco D. Mazzitelli}
\affiliation{Instituto Balseiro, Centro Atómico Bariloche, San Carlos de Bariloche, Argentina}


\begin{abstract}

We study how the Lamb shift of a static atom is modified when a nearby planar
body rotates rigidly about its normal while the atom is held at a fixed
distance $a$. We derive a general formula for the shift in terms of the
angularly Doppler-shifted reflection coefficients of the surface, valid for any
axially symmetric planar material. Expanding the result to second order in the
angular velocity $\Omega$, we identify two independent contributions associated
with the orbital and spin components of the electromagnetic angular momentum.
The orbital contribution, proportional to $(\Omega\rho)^2$, reproduces locally
the Lamb shift induced by a surface translating at the tangential velocity
$\Omega\rho$, whereas the spin contribution, proportional to $(a\Omega)^2$,
originates from the rotational Doppler shift of the photon helicity and
survives even on the rotation axis. We first illustrate the formalism using a
graphene sheet and then apply it to finite-thickness Drude and plasma
conductors and to doped semiconductors. Rotation enhances the Casimir--Polder
interaction for graphene and metallic surfaces, whereas it weakens it for doped
semiconductors, depending on whether the carrier plasma frequency reaches the
near-field scale $1/a$. Above a threshold angular velocity, the atomic level
also acquires a finite linewidth, providing a spectroscopic signature of
quantum friction. Furthermore, rotation induces a novel component of the Casimir–Polder force, which is perpendicular to both the standard normal attraction and the tangential quantum-friction force.

\end{abstract}

\maketitle
\section{Introduction}
\label{sec:intro}

An atom placed near a material body has its energy levels shifted: the
virtual photons that dress the atom can be reflected by the body before
being reabsorbed, and the radiative self-energy of each level acquires a
position-dependent part. This correction to the ground-state energy is the
origin of the celebrated Casimir-Polder attraction \cite{Casimir-Polder}.

In this paper we study what happens to this shift when the body rotates
rigidly at angular velocity $\Omega$ about its normal, the atom remaining
static. The body is a planar surface characterized entirely by its
electromagnetic reflection coefficients; the rotation enters only
kinematically, through an angular Doppler shift of those coefficients. We
use a graphene sheet to introduce the method and to derive the general
formula for the shift, and then apply that same formula to other media,
finite-thickness conducting and semiconducting disks among them.

This problem belongs to the family of fluctuation-induced effects between
bodies in relative motion. For two graphene sheets in relative sliding
motion, quantum friction has the distinctive feature of a velocity
threshold: dissipation is absent unless the relative speed exceeds the Fermi
velocity $\vF$ of the Dirac quasiparticles~\cite{friction} (for background
on Casimir friction and on graphene see Refs.~\cite{Volokitin,Geim,Woods}). 
Rotational vacuum friction has been studied for small spinning particles, whose 
fluctuating anisotropic polarizability radiates and dissipates at any
$\Omega$~\cite{ManjavacasGdA,ZhaoPendry}. The configuration considered here
is complementary: the rotating body is an extended, axially symmetric
sheet, and the probe is a static, pointlike atom. The influence of a moving
medium on a nearby atom is also being pursued experimentally: it has been
proposed to detect motion-induced (quantum-friction) effects through the
velocity dependence of the geometric phase and decoherence of a
nitrogen-vacancy center held above a rotating disk coated with n-doped
silicon or gold~\cite{DecaLombardo}, a setup geometrically close to the one
analyzed here. The rotating sheet sees
every electromagnetic mode of angular momentum $m$ about the axis at the
Doppler-shifted frequency $\omega-m\Omega$, and, as we shall see, the atom
senses this frequency reshuffling in two distinct ways: through the orbital
angular momentum of the exchanged photons, available only off the axis,
which reproduces locally the physics of a sheet sliding at the velocity
$v=\Omega\rho$ of the material beneath the atom; and through the photon
helicity, a rotational Doppler shift of the polarization that survives even
on the axis.

The paper is organized as follows. In Sec.~\ref{sec:sheet} we construct
the response of the rotating sheet, characterized by its reflection
coefficients, and identify the angular Doppler shift; in
Sec.~\ref{sec:static} we express the level shift in terms of the
reflection coefficients of the static sheet, and in
Sec.~\ref{sec:channels} we convert the problem to angular-momentum
channels and present the general formula for the shift.
Section~\ref{sec:smallOmega} contains the small-$\Omega$ expansion,
numerical estimates in the retarded regime, and the dissipative threshold
above which the level acquires a width. Section~\ref{sec:drude} applies the
general formula to further media, finite-thickness Drude and plasma
conductors and doped semiconductors, treating each as an example of the
same construction. Section~\ref{sec:concl} presents our conclusions.
Appendix~\ref{app:scattering} solves explicitly the Maxwell equations for
the field of the point dipole in the presence of the rotating sheet and derives
the general formula from that solution; Appendix~\ref{app:functional} gives an
independent, functional-integral derivation of the same formula, and
Appendix~\ref{app:bulkderiv} the closed form of the $O(\Omega^2)$ coefficients
for bulk media.

\begin{figure}[t]
  \centering
  \begin{tikzpicture}[>={Latex[length=1.6mm]},scale=1.0]
    \begin{scope}[xshift=0cm]
      \def\rx{2.0}\def\ry{0.6}\def\gap{1.7}
      \draw[thick,fill=gray!12] (0,0) ellipse ({\rx} and {\ry});
      \draw[densely dashed] (0,-\ry-0.5) -- (0,\gap+0.9);
      \draw[->,thick,blue!70!black] (0.95,0) arc (0:305:0.95 and 0.285);
      \node[blue!70!black] at (1.38,0.40) {$\Omega$};
      \fill[red!70!black] (1.2,\gap) circle (0.06);
      \node[above] at (1.2,{\gap+0.08}) {\small atom};
      \draw[densely dotted] (0,\gap) -- (1.2,\gap);
      \node[above] at (0.55,\gap) {\small $\rho$};
      \draw[<->] (1.75,0) -- (1.75,\gap);
      \node[right] at (1.78,\gap/2) {$a$};
      \node[below] at (0,-\ry-0.55) {\small axis};
      \node[left] at (-\rx-0.05,0) {\small sheet};
    \end{scope}
    \begin{scope}[xshift=6.8cm,yshift=0.55cm]
      \def\Ro{1.75}\def\Rc{0.9}
      \fill[orange!22,even odd rule] (0,0) circle (\Ro) (0,0) circle (\Rc);
      \draw[thick] (0,0) circle (\Ro);
      \draw[densely dashed] (0,0) circle (\Rc);
      \draw[->] (0,0) -- (145:\Ro);
      \node at (145:{\Ro+0.22}) {$\Rdisk$};
      \draw[->] (0,0) -- (-90:\Rc);
      \node[right] at (-90:0.46) {$r_c$};
      \fill (0,0) circle (0.03);
      \fill[red!70!black] (25:1.3) circle (0.06);
      \node[above right] at (25:1.32) {\small atom};
      \draw[->,thick,blue!70!black] (5:{\Ro+0.3}) arc (5:75:{\Ro+0.3});
      \node[blue!70!black] at (40:{\Ro+0.55}) {$\Omega$};
    \end{scope}
  \end{tikzpicture}
  \caption{Left: a static atom at height $a$ above the rotating sheet, at
  lateral distance $\rho$ from the rotation axis. Right: face-on view. For
  $\rho<r_c=\vF/\Omega$ the local speed $\Omega\rho$ of the material beneath
  the atom is below the Fermi velocity and the level shift is strictly real;
  for an atom above the annulus $\rho>r_c$ (shaded), and for sufficiently
  low transition frequencies, the level also acquires an $\Omega$-induced
    width [Eq.~\eqref{eq:threshold}].} 
\label{fig:geom}
\end{figure}
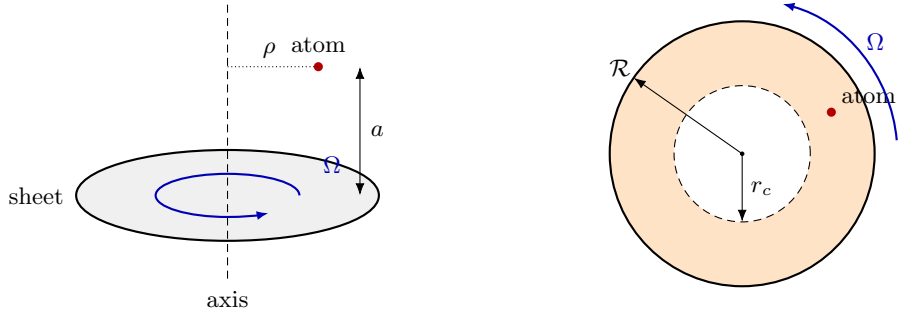

Let us now make explicit the ingredients of the system we consider and of the
description and approximations we use in order to analyze it. 
We first clarify our conventions: regarding the space-time metric, we use
Euclidean signature, $\hbar=c=1$, $\kpar=(k_0,k_1,k_2)$ for the frequency and
in-plane momenta, $\kq=(k_1,k_2)$ and $q=|\kq|$, as in Ref.~\cite{friction}. The
material plane occupies the plane $x_3=0$; the atom sits at
$\rA=(\rho,\phi_A,a)$ in cylindrical coordinates (Fig.~\ref{fig:geom}), $a$
denoting throughout the atom-sheet distance and $\rho$ the distance to the
rotation axis. We use Heaviside-Lorentz units, in which $\mathbf d=\alpha\mathbf
E$ defines $\alpha$ with $\alpha_{\rm HL}=4\pi\,\alpha_{\rm Gauss}$.

Let us also state, at the outset, the assumptions that determine the domain of
validity of our treatment: the atom's coupling to the electromagnetic field is
described in the dipole approximation; moreover,  it is assumed  to be in its
ground state, and to  have isotropic dynamic polarizability $\alpha(\omega)$
(which appears at second order in the dipole coupling). 

For the graphene example, the only input from the medium side is the one-loop
vacuum polarization tensor (VPT) of undoped, gapless graphene. The photon
propagator is dressed by VPT insertions and nothing else (no vertex corrections,
no fermion self-energies), and the response of the rotating sheet is assumed to
be local and in equilibrium in the comoving frame. Finite temperature, doping or
gapping of the graphene sheet, disorder, strain and nonlocal corrections beyond
one loop are not considered. On the other hand, the conducting and
semiconducting disks of Sec.~\ref{sec:drude} are instead described by a local
bulk dielectric function. 

\section{The rotating medium}
\label{sec:sheet}

Our derivation of the general formula describing the phenomenon we study
begins from a definite example: a graphene sheet, whose response is known in
closed form; the only graphene-specific input is its vacuum polarization tensor
(VPT). 
Everything downstream shall be  written in terms of reflection
coefficients, so the passage to other media will amount to just changing that
input.

In order to analyze the response of the rotating graphene sheet, we write
the Euclidean action of the electromagnetic field coupled to the Dirac
quasiparticles confined to the sheet:
\begin{equation}
  S[A;\bar\psi,\psi]=S^{(0)}_g[A]+S^{(0)}_d[\bar\psi,\psi]
  +S^{(\mathrm{int})}_{dg}[\bar\psi,\psi,A]\,,
  \label{eq:action}
\end{equation}
with $S^{(0)}_g[A]=\tfrac14\!\int d^4x\,F_{\mu\nu}F_{\mu\nu}$ and the matter
terms localized on the plane. Integrating out the fermions produces, to
quadratic order in the coupling, a surface term governed by 
$\Pi_{\alpha\beta}$, the VPT of the sheet. 

When the sheet is at rest, the VPT is invariant under time translations,
rotations and translations on the plane, so the Ward identity
$k_\alpha\widetilde\Pi_{\alpha\beta}=0$ fixes its form in terms of two
projectors,
\begin{equation}
  \widetilde\Pi_{\alpha\beta}(\kpar)
  =g_t(\kpar)\,\Pt_{\alpha\beta}+g_l(\kpar)\,\Pl_{\alpha\beta}\,,
  \label{eq:VPTdecomp}
\end{equation}
with $\Pt$, $\Pl$ (transverse and longitudinal with respect to the in-plane
momentum) built from $\delta_{\alpha\beta}$, $k_\alpha$ and
$n_\alpha=(1,0,0)$. For gapless graphene, with $N$ two-component flavors
($N=4$ for monolayer graphene) and $\aN\equiv e^2N/16$,
\begin{align}
  g_t(\kpar) & =\aN\sqrt{k_0^2+\vF^2 k_q^2}\,,
  \label{eq:gt}\\
  g_l(\kpar) & =\aN\,\frac{k_0^2+k_q^2}{\sqrt{k_0^2+\vF^2 k_q^2}}\,,
  \label{eq:gl}
\end{align}
the one-loop kernels of the massless $2+1$ Dirac
theory~\cite{friction,BFGV}, adequate for the momenta $q\sim1/a$ of
interest, far below the lattice scale. 

It is convenient to trade the VPT for reflection coefficients. A single VPT
insertion in the photon line is the Born approximation to the reflection,
$r_s\simeq g_s/(2K)$ per polarization, with $K=\sqrt{k_0^2+q^2}$; repeated
scatterings form a geometric series 
which resums into the Fresnel-like (Lifshitz) coefficients for the transverse-magnetic (TM) and transverse-electric (TE) polarizations

\begin{equation}
  r_{\rm tm}=\frac{g_l}{g_l+2K}\equiv R_l\,,\qquad
  r_{\rm te}=-\,\frac{g_t}{g_t+2K}\equiv -R_t\,,
  \label{eq:refl}
\end{equation}
in agreement with the graphene literature~\cite{FMV}; both $R_{l,t}$ are
positive for a passive sheet. One feature of \eqref{eq:gl} will dominate
everything that follows. At zero Euclidean frequency \eqref{eq:gl} gives
$g_l(0,q)=\aN q/\vF$, so
\begin{equation}
  R_l(0,q)=\frac{\aN}{\aN+2\vF}\,,
  \label{eq:Rl0}
\end{equation}
independent of $q$. 

Consider now the sheet rotating rigidly at angular velocity $\Omega$ about
$x_3$. The material response is local and in equilibrium in the comoving
frame, where the VPT is the static kernel
\eqref{eq:VPTdecomp}-\eqref{eq:gl}; within this approximation, the effect
of the rotation is purely kinematic, and the derivation reduces to
identifying the comoving frequency
seen by a laboratory mode. The comoving frame is the corotating one,
$\bar\phi=\phi-\Omega t$, $\bar r=r$, $\bar t=t$, with the operator
identity
\begin{equation}
  \partial_t\big|_{\phi}=\partial_{\bar t}-\Omega\,\partial_{\bar\phi}\,,
  \label{eq:dt}
\end{equation}
the angular analog of the Galilean $\partial_t=\partial_{\bar
t}-v\,\partial_{\bar x_1}$. 

Note that, since the configuration is invariant under simultaneous time
translations and rotations about $x_3$, the conserved labels are the
frequency $\omega$ and the angular momentum $m$, and the response
block-diagonalizes in $(\omega,m)$. Indeed, a laboratory mode $e^{-i\omega
t+im\phi}$, when rewritten in corotating coordinates, becomes
$e^{-i(\omega-m\Omega)t+im\bar\phi}$: the comoving frequency is
$\bar\omega=\omega-m\Omega$, while $m$ and the in-plane momentum modulus $q$,
invariant under a planar rotation, are unchanged. For a field that carries
indices, such as the gauge field, the corotating map rotates not only the
argument of the mode but also its in-plane components, and the label $m$ for
which the shift holds is the \emph{total} angular momentum of the mode, orbital
plus polarization. Since the projectors in \eqref{eq:VPTdecomp} transform 
covariantly, the shift is carried entirely by the scalar functions. 
Indeed, in the basis of frequency, total angular momentum $m$ and in-plane modulus $q$,
continued to Euclidean frequency,
\begin{equation}
  R_s\;\longrightarrow\;R_s(\bar k_0,q)\,,\qquad
  \bar k_0=k_0+i\,m\,\Omega
  \quad\Longleftrightarrow\quad \bar\omega=\omega-m\Omega\,,
  \label{eq:shift}
\end{equation}
with the static coefficients at the shifted frequency. The photon kinematic
factors ($K$ and the propagation factors below) are defined in the laboratory
frame and are untouched. The shift also has a local reading, exploited later: at
radius $r$ the azimuthal wavenumber of the $m$-th channel is $m/r$, so
$m\Omega=(\Omega r)(m/r)$ is the linear Doppler shift produced by the local
velocity $\Omega r$ acting on the azimuthal momentum component. The shift
$\omega\to\omega-m\Omega$ of the angular-momentum channels is the general
signature of a body rotating about the axis of the channels: it appears in the
low-velocity electrodynamics of rotating conductors and
dielectrics~\cite{VanBladel,Shiozawa}, in the reflection of cylindrical waves by
a rotating cylinder, where it leads to superradiant amplification for
$\omega<m\Omega$~\cite{Zeldovich,BekensteinSchiffer,RotExp}, in the explicit
solutions of the scattering problem for rotating spheres and
cylinders~\cite{DeZutter}, and in the scattering-theory description of the
quantum radiation of rotating objects~\cite{MaghrebiRot}; for a rotating light
beam it is the rotational frequency shift~\cite{BialynickiBirula}. In
Appendix~\ref{app:scattering} we solve explicitly the Maxwell equations for the
field of the atomic dipole in the presence of the rotating sheet, along these
lines, and derive the general formula of Sec.~\ref{sec:channels} from that
solution.

We now argue that the same conclusion for the coefficients of rotating graphene
holds when considering a microscopic description: writing the massless $2+1$
Dirac theory of the quasiparticles in corotating coordinates, the time evolution
is generated by
\begin{equation}
  i\,\partial_t\psi=\big(H_0-\Omega\,J_z\big)\psi\,,
  \qquad J_z=-i\,\partial_{\bar\phi}+\half\,\sigma_3\,,
  \label{eq:HJz}
\end{equation}
with $H_0$ the static graphene Hamiltonian: the rotation enters as a
chemical potential conjugate to the total angular momentum, as is familiar
from quantum field theory in rotating frames~\cite{Vilenkin,Iyer}. The
corotating Hamiltonian shares its eigenfunctions with $H_0$, with shifted
eigenvalues $\bar E=E-j\Omega$, $j$ being the conserved total angular
momentum of the quasiparticle mode; evaluating the current correlators with
\eqref{eq:HJz} and writing $\phi-\phi'=(\bar\phi-\bar\phi')+\Omega(t-t')$
in their channel expansion shows that the laboratory channel $(\omega,m)$
is the corotating channel $(\omega-m\Omega,m)$, which is \eqref{eq:shift}
again. The spectral content matters for dissipation: for quasiparticles
confined to a disk of radius $\Rdisk$, a state of angular momentum $j$ has
a classical turning point at $r_t=\vF|j|/E$, so $r_t\le\Rdisk$ enforces
$E\ge\vF|j|/\Rdisk$ and
\begin{equation}
  \bar E\;=\;E-j\,\Omega\;\ge\;|j|\,\Big(\frac{\vF}{\Rdisk}-\Omega\Big)\,,
  \label{eq:bound}
\end{equation}
strictly positive for every excitation if and only if $\Omega\Rdisk<\vF$.
Below this threshold the rotating ground state is the adiabatically
continued Dirac sea and no dissipation can occur. Above it, modes with
turning points at $r>r_c\equiv\vF/\Omega$ acquire negative corotating
energies ($E>0$ with $\bar E<0$): exciting a pair out of these superradiant
modes lowers the corotating energy, the rotation acting as the energy
reservoir.

Equation
\eqref{eq:bound} is a \emph{global} statement about the disk:
$\Omega\Rdisk<\vF$ is the condition that its adiabatic corotating vacuum is
stable as a whole; the local reading of the threshold, the one probed by the atom at a given radius, is discussed in Sec. VB.

The replacement \eqref{eq:shift} carries the entire effect of the rotation,
and its status differs for a sheet and for a bulk body. For graphene it is
exact at one loop for the kernel in the corotating frame: \eqref{eq:HJz}
shows that each angular-momentum channel of the vacuum polarization is
rigidly shifted, $\Pi_m(\omega)\to\Pi_m(\omega-m\Omega)$, with no further
structure (the passage to the laboratory frame is discussed in
Appendix~\ref{app:scattering}).

\section{The level shift for \texorpdfstring{$\Omega = 0$}{Omega = 0}}
\label{sec:static}

Let $|g\rangle$ denote the atomic ground state and $\{|e\rangle\}$ its
excited states, with excitation energies $\omega_{eg}\equiv E_e-E_g>0$ and
dipole matrix elements $\mathbf d_{eg}\equiv\langle e|\hat{\mathbf
d}|g\rangle$, $\hat{\mathbf d}$ being the electric dipole operator. The
atom couples to the quantized field through $-\hat{\mathbf
d}\cdot\mathbf E(\rA)$, and second-order perturbation theory gives, for
the correction to the ground-state energy,
\begin{equation}
  \delta E=-\sum_{e}\sum_{\lambda}
  \frac{\big|\mathbf d_{eg}\cdot\mathbf E_{\lambda}(\rA)\big|^2}
  {\omega_{eg}+\omega_{\lambda}}\,,
  \label{eq:PT2}
\end{equation}
a sum over virtual atomic excitations and over the modes $\lambda$ of the
field in the presence of the sheet, with frequency $\omega_\lambda$ and
electric mode function $\mathbf E_\lambda(\rA)$. For a static configuration
$\delta E$ is real; this may change for a rotating sheet, as we shall see.

The standard manipulation (equivalently, the in-out effective action with the
atom as a localized polarizable insertion, whose real part is the shift and
whose imaginary part the decay probability) trades the double sum for a single
integral over imaginary frequencies,
\begin{equation}
  \delta E=-\int_{-\infty}^{\infty}\frac{dk_0}{2\pi}\;\alpha(ik_0)\;
  \mathrm{Tr}\,\mathcal E(ik_0;\rA,\rA)\,,
  \label{eq:WS}
\end{equation}
where
\begin{equation}
  \alpha(ik_0)=\frac{2}{3}\sum_e
  \frac{\omega_{eg}\,|\mathbf d_{eg}|^2}{\omega_{eg}^2+k_0^2}
  \label{eq:alpha}
\end{equation}
is the isotropic dynamic polarizability on the imaginary axis, smooth and
free of resonant denominators, and $\mathcal E_{ij}(ik_0;\mathbf r,\mathbf
r')$ is the Euclidean correlator of the electric field. Only the
scattering part of $\mathcal E$, the part involving the sheet, is kept: the
free part reproduces the position-independent free-space Lamb shift.

The scattering part describes a virtual photon emitted by the atom,
reflected by the sheet, and reabsorbed by the atom. In the mixed representation
(frequency $k_0$, in-plane momentum $\kq$, position $x_3$) the free
Euclidean propagator between the planes is $e^{-aK}/(2K)$: every Euclidean
mode decays exponentially in $x_3$ (in real frequencies, the modes with
$\omega<q$ are the evanescent near fields, dominant at the separations of
interest), so the round trip contributes $e^{-2aK}$, the sheet one
reflection coefficient per polarization, and the polarization vectors the
tensor structure. We write
\begin{equation}
  \mathcal E^{\rm sc}_{ij}(ik_0;\rA,\rA)=\int\frac{d^2\kq}{(2\pi)^2}\;
  \frac{e^{-2aK}}{4K}\;M_{ij}(k_0,\kq)\,,
  \label{eq:Esc}
\end{equation}
the overall normalization being fixed by the perfect-mirror limit below,
and determine $M$ per polarization. For TE (s) waves the electric field
points along $\hat e=\hat z\times\hat k$ for both legs and carries one
factor of frequency per leg ($\mathbf E=-\partial_t\mathbf A$); the dyadic
is $(\omega^2/c^2)\,r_{\rm te}\,\hat e_i\hat e_j$, and the Euclidean
continuation $\omega^2\to-k_0^2$, together with $r_{\rm te}=-R_t$, gives
\begin{equation}
  M^{({\rm te})}_{ij}=k_0^2\,R_t\,\hat e_i\hat e_j\,,
  \label{eq:Mte}
\end{equation}
the two minus signs compensating; this is why the TE channel also
contributes attractively for a mirror. For TM (p) waves the electric field
lies in the plane of incidence and must be orthogonal to the wavevector,
which differs for the two legs, $k^{\mp}=(\kq,\mp k_z)$ with
$k_z=\sqrt{\omega^2-q^2}$: the unit vectors are
$\hat e_p^{\,\mp}=(\pm k_z\,\hat k+q\,\hat z)\,c/\omega$. The correlator
at coincident points is symmetric in the indices, so the two orderings of
the legs are summed,
\begin{equation}
  \half\big(\hat e_p^{\,+}\otimes\hat e_p^{\,-}
  +\hat e_p^{\,-}\otimes\hat e_p^{\,+}\big)
  =\big(-k_z^2\,\hat k\hat k+q^2\,\hat z\hat z\big)\,\frac{c^2}{\omega^2}\,,
\end{equation}
the mixed $\hat z\hat k$ terms, antisymmetric, canceling. The
$\omega^2/c^2$ from the two electric-field vertices cancels the
normalization of the polarization vectors, and the continuation
$k_z^2\to-K^2$ gives, with $r_{\rm tm}=R_l$,
\begin{equation}
  M=R_l\,\big(K^2\,\hat k\hat k+q^2\,\hat z\hat z\big)
  +R_t\,k_0^2\,\hat e\hat e\,.
  \label{eq:M}
\end{equation}
Taking the trace in \eqref{eq:Esc} and inserting into \eqref{eq:WS}, the
shift produced by the static sheet is
\begin{equation}
  \delta E^{(0)}(a)=-\int_{-\infty}^{\infty}\frac{dk_0}{2\pi}
  \int_0^{\infty}\frac{q\,dq}{2\pi}\;\alpha(ik_0)\,
  \frac{e^{-2aK}}{4K}\,
  \Big[(2K^2-k_0^2)\,R_l+k_0^2\,R_t\Big]\,,
  \label{eq:CP0}
\end{equation}
since $(K^2+q^2)=2K^2-k_0^2$, in agreement with Wylie and
Sipe~\cite{WylieSipe}. In the
perfect-mirror limit $R_l,R_t\to1$ and, at large $a$ and with
$\alpha(ik_0)\to\alpha(0)$, Eq.~\eqref{eq:CP0} gives $\delta
E=-3\,\alpha(0)/(32\pi^2a^4)$, the Casimir-Polder result in these units.

\section{The level shift for \texorpdfstring{$\Omega \neq 0$}{Omega neq 0}}
\label{sec:channels}
With the sheet at rest the natural photon labels are $(k_0,\kq)$, and
\eqref{eq:CP0} is the whole story. A rotating sheet destroys in-plane
translation invariance but preserves time translations and rotations about
the axis: the good quantum numbers are $(k_0,m)$, the basis in which the
response \eqref{eq:shift} is diagonal, so the photon modes must be
rewritten in it. The conversion is provided by the Jacobi-Anger expansion,
\begin{equation}
  e^{i\kq\cdot\boldsymbol\rho}
  =\sum_{m=-\infty}^{\infty} i^{m}\,J_m(q\rho)\,
  e^{im(\phi_\rho-\phi_k)}\,,
  \label{eq:JA}
\end{equation}
$\phi_k$ being the direction of $\kq$: a photon of in-plane momentum $q$,
observed at lateral distance $\rho$ from the axis, is found in the
angular-momentum channel $m$ with amplitude $J_m(q\rho)$, and the
completeness relation $\sum_mJ_m^2(x)=1$ makes the weights $J_m^2(q\rho)$ a
normalized probability distribution over channels. Their shape has a
classical reading, used repeatedly below: a photon of in-plane momentum $q$
crossing the atom's location at angle $\theta$ with the radial direction
carries orbital angular momentum
\begin{equation}
  \ell_z=\rho\,q\sin\theta
  \label{eq:classical}
\end{equation}
(momentum times impact parameter). Accordingly, $J_m(q\rho)$ is
exponentially small in the classically forbidden region $|m|>q\rho$, and
for $q\rho\gg1$ the weights approach the classical distribution of
$\ell_z=q\rho\sin\theta$ with $\theta$ uniform,
\begin{equation}
  J_m^2(q\rho)\;\simeq\;\frac{1}{\pi\sqrt{(q\rho)^2-m^2}}\,,\qquad
  |m|<q\rho\,,
  \label{eq:Debye}
\end{equation}
the oscillation-averaged Debye asymptotics. Near the axis the opposite
happens: $J_m(0)=\delta_{m0}$, and a point on the axis communicates only
with the lowest channels.

For a vector field one more element enters: a rotation about the axis acts
both on the mode's argument and on its in-plane components, so the angular
momentum the rotating sheet couples to is the total one, orbital plus the
helicity carried by the polarization. In circular components
$E_\pm=(E_1\pm iE_2)/\sqrt2$, which pick up phases $e^{\mp i\beta}$ under a
rotation by $\beta$, single-valuedness forces, for a mode of total angular
momentum $m$,
\begin{equation}
  E_z\;\propto\;J_{m}(q\rho)\,e^{im\phi}\,,\qquad
  E_\pm\;\propto\;J_{m\pm1}(q\rho)\,e^{i(m\pm1)\phi}\,:
  \label{eq:modes}
\end{equation}
the $z$ component is a scalar under in-plane rotations and carries the full
$m$ as orbital angular momentum, while each circular component carries one
unit in its polarization and $m\pm1$ in its argument. (Equivalently, the
circular components of any unit vector attached to the wave, such as $\hat
k$, carry factors $e^{\mp i\phi_k}$ that shift the Bessel order in
\eqref{eq:JA} by one unit.) On the axis, since
$J_n(0)=\delta_{n0}$, the only modes with nonvanishing electric field are
$m=0$ (through $E_z$) and $m=\pm1$ (through $E_\pm$): a point on the axis
cannot exchange orbital angular momentum with the sheet, there being no
lever arm, but it can still exchange the photon's intrinsic unit.

It remains to decompose the dyadic \eqref{eq:M} accordingly. The
$\hat z\hat z$ piece is a scalar and goes over into $q^2J_m^2(q\rho)R_l$
per channel. For the in-plane part, since
$\hat k_a\hat k_b+\hat e_a\hat e_b=\delta_{ab}$ while
$\hat k_a\hat k_b-\hat e_a\hat e_b$ has only $e^{\pm2i\phi_k}$ Fourier
components,
\begin{equation}
  K^2R_l\,\hat k_a\hat k_b+k_0^2R_t\,\hat e_a\hat e_b
  =\half\big(K^2R_l+k_0^2R_t\big)\,\delta_{ab}
  +\half\big(K^2R_l-k_0^2R_t\big)
  \big(\hat k_a\hat k_b-\hat e_a\hat e_b\big)\,.
  \label{eq:helsplit}
\end{equation}
The first piece is helicity diagonal, each circular component attaching the
weight $J_{m\mp1}^2(q\rho)$ by \eqref{eq:modes}; the second (quadrupole)
piece is traceless, connects the channels $m$ and $m\mp2$, and does not
contribute for an isotropic atom. It would matter for atoms with anisotropic
polarizabilities, or for a parity-breaking (gyrotropic) surface: there the
$m\leftrightarrow m\mp2$ coupling no longer pairs $\pm m$ symmetrically, and
a term \emph{linear} in $\Omega$ can survive the channel sum, in contrast to
the strictly even-in-$\Omega$ shift found here. We do not pursue this case.
Evaluating the reflection
functions of each total-angular-momentum channel at the Doppler-shifted
argument \eqref{eq:shift}, with the kinematic factors $K$, $q$, $k_0$
untouched, the level shift of the atom is
\begin{align}
  \delta E(\rho;\Omega)=-\sum_{m=-\infty}^{\infty}
  \int_{-\infty}^{\infty}\frac{dk_0}{2\pi}
  \int_0^{\infty}\frac{q\,dq}{2\pi}\;\alpha(ik_0)\,
  \frac{e^{-2aK}}{4K}\,
  \Big\{ & q^2\,J_m^2(q\rho)\,R_l(\bar k_0,q)\nonumber\\[-2pt]
  +\,\half\,\big[J_{m-1}^2(q\rho)+J_{m+1}^2(q\rho)\big]\,
  & \big[K^2R_l+k_0^2R_t\big](\bar k_0,q)\Big\}\,,
  \label{eq:masterf}
\end{align}
with $\bar k_0=k_0+im\Omega$; the square bracket indicates that the
reflection functions, and only they, carry the shifted argument. The
explicit solution of the scattering problem behind this formula, the Maxwell
equations for the field of the dipole in the presence of the rotating sheet
solved in cylindrical partial waves, is given in Appendix~\ref{app:scattering},
where \eqref{eq:masterf} is obtained from the field reflected back to the
position of the atom. An independent derivation, by a
first-order functional-integral computation in which the free photon
correlator is written directly in the cylindrical basis, is given in
Appendix~\ref{app:functional}.

A word on the analytic continuation, since for $\Omega\neq0$ the reflection
functions are sampled off the Euclidean axis at $\bar k_0=k_0+im\Omega$.
The $R_s(k_0,q)$ are the responses continued from the retarded ones,
analytic in the upper half complex-$\omega$ plane and, on the imaginary
axis, even and real for a passive medium. The shifted argument
$\bar k_0=k_0+im\Omega$ displaces the contour parallel to the imaginary
axis by $m\Omega$; \eqref{eq:masterf} is the correct continuation as long as
this strip contains no singularity of $R_s$, which holds for the dissipative
Drude response (poles at $k_0=0,-\gamma$ on the negative imaginary axis,
never crossed) and for graphene. Where a real-frequency branch point or
surface-mode pole is reached, the contour acquires a discontinuity; that is
precisely the imaginary part computed in Sec.~\ref{ssec:dissipation}. Thus
the real series below is generated by the smooth continuation, and the
dissipative onset by the contour crossing the matter cut.

Two checks. For $\Omega=0$ nothing depends on $m$, the completeness sum
collapses each weight, and \eqref{eq:masterf} reduces to the planar result
\eqref{eq:CP0}, independently of $\rho$: a static infinite sheet cannot
know where the axis is. On the axis, $J_n(0)=\delta_{n0}$ keeps only $m=0$
in the $E_z$ piece and $m=\pm1$ in the circular pieces:
\begin{equation}
  \delta E_{\rm axis}(\Omega)=-\int_{-\infty}^{\infty}\frac{dk_0}{2\pi}
  \int_0^{\infty}\frac{q\,dq}{2\pi}\,\alpha(ik_0)\,\frac{e^{-2aK}}{4K}
  \Big[q^2R_l(k_0,q)+\half\sum_{\pm}
  \big[K^2R_l+k_0^2R_t\big](k_0\pm i\Omega,\,q)\Big]\,.
  \label{eq:axis}
\end{equation}
On the axis the rotation is felt only through the photon helicity: the
effective Doppler shift is $\pm\Omega$, one unit per photon, rather than
$m\Omega$ with $m\sim q\rho$. This $\pm\Omega$ shift of circularly
polarized light reflected off a rotating body is the rotational Doppler
effect of optics~\cite{Garetz,Courtial}, and it is the planar counterpart
of the relative-rotation physics of a spinning particle near a
surface~\cite{ZhaoPendry,ManjavacasGdA}: here it is the sheet, rather than
the atom, that rotates. A scalar probe on the axis would see no $\Omega$
dependence at all, to all orders in the VPT, since every scalar mode
attached to an axial point has $m=0$.

\section{The shift at small \texorpdfstring{$\Omega$}{Omega}}
\label{sec:smallOmega}

For an isotropic atom the shift is even in $\Omega$: reversing the sense of
rotation is an in-plane reflection, under which \eqref{eq:masterf} is
invariant (relabel $m\to-m$); equivalently, the VPT of gapless, undoped
graphene is parity even (no Chern-Simons term), so in the present
isotropic setup no term odd in $\Omega$ can appear. The leading dependence is therefore $O(\Omega^2)$, and it can
be computed in closed form for all $\rho$, because the required moments of
the Bessel weights are known exactly. Applying Parseval's theorem to the
Jacobi-Anger expansion $e^{ix\sin\theta}=\sum_m J_m(x)\,e^{im\theta}$,
i.e.\ integrating $|e^{ix\sin\theta}|^2$ and $|\partial_\theta
e^{ix\sin\theta}|^2$ over $\theta$ and using $J_{-m}=(-1)^mJ_m$, gives
\begin{equation}
  \sum_m J_m^2(x)=1\,,\quad \sum_m m\,J_m^2(x)=0\,,\quad
  \sum_m m^2 J_m^2(x)=\frac{x^2}{2}\,,\quad
  \sum_m m^2 J_{m\mp1}^2(x)=\frac{x^2}{2}+1\,.
  \label{eq:sumrules}
\end{equation}
The third is the classical mean square of \eqref{eq:classical},
$\langle(q\rho\sin\theta)^2\rangle_\theta=(q\rho)^2/2$, here exact for all
$x$; the fourth adds the photon's intrinsic unit. Expanding
$R(k_0+im\Omega)=R+im\Omega\,\partial_{k_0}R
-\half m^2\Omega^2\,\partial^2_{k_0}R+\dots$ in \eqref{eq:masterf}, the
linear terms drop by the second sum rule and the quadratic ones are fixed
by the third and fourth:
\begin{align}
  \delta E(\rho;\Omega)-\delta E^{(0)}
  =\;&\frac{(\Omega\rho)^2}{4}\int_{-\infty}^{\infty}\frac{dk_0}{2\pi}
  \int_0^{\infty}\frac{q\,dq}{2\pi}\,\alpha(ik_0)\,
  \frac{q^2\,e^{-2aK}}{4K}\,
  \Big[(2K^2-k_0^2)\,\partial^2_{k_0}R_l+k_0^2\,\partial^2_{k_0}R_t\Big]
  \nonumber\\[2pt]
  +\;&\frac{\Omega^2}{2}\int_{-\infty}^{\infty}\frac{dk_0}{2\pi}
  \int_0^{\infty}\frac{q\,dq}{2\pi}\,\alpha(ik_0)\,
  \frac{e^{-2aK}}{4K}\,
  \Big[K^2\,\partial^2_{k_0}R_l+k_0^2\,\partial^2_{k_0}R_t\Big]
  \;+\;O(\Omega^4)\,.
  \label{eq:Omega2}
\end{align}
The two terms have distinct meanings; in both, the second derivatives
$\partial_{k_0}^2 R_{l,t}$ are evaluated at the \emph{unshifted} argument
$k_0$ on the retarded response continued to the imaginary axis, the
$O(\Omega^2)$ coefficients of the Doppler average
$\tfrac12[R(k_0+im\Omega)+R(k_0-im\Omega)]$. The orbital term,
$\propto(\Omega\rho)^2$, is the $O(v^2)$ expansion of the shift produced by
a sheet whose response is Doppler shifted by the local velocity
$v=\Omega\rho$ beneath the atom: the classical identification
$m\Omega=(\Omega\rho)(q\sin\theta)$ becomes exact at this order because
only the second moment of $m$ enters, and that moment coincides with the
classical one. It is the local-density (sliding) result, valid even at
$\rho\lesssim a$, and it vanishes on the axis. The spin term,
$\propto(a\Omega)^2$ once the $a$ scaling of the integral is extracted, is
the rotational Doppler shift of the polarization: every exchanged photon,
wherever the atom sits, has its circular components shifted by $\mp\Omega$.
It is independent of $\rho$ and is the only survivor on the axis, where it
reproduces the expansion of \eqref{eq:axis}.

This decomposition matches the structure of the derivative expansion of
interaction functionals of slowly varying fields~\cite{DE}, applied to the
velocity field $v_i=\Omega\,\epsilon_{ij}x_j$ of the sheet: the leading
order is the local-density value at $v(\rho)=\Omega\rho$, and the
corrections are classified by gradient invariants, each derivative
accompanied by one power of $a$. For rigid rotation the strain rate and the
divergence vanish, the vorticity is forbidden by parity, and the first
corrections are $O\big((a\Omega)^2\big)$. The spin term shares that scaling
but is not itself a velocity-gradient correction: it is the rotational
Doppler shift of the photon helicity, a spin-connection effect attached to
the polarization rather than to the local strain of $v_i$, and the
derivative-expansion counting serves only to fix its $(a\Omega)^2$ size. The
ratio of the two terms, $(a/\rho)^2$, shows that the local approximation is
protected to second order in $a/\rho$. The expansion
\eqref{eq:Omega2} holds for $\Omega\rho\ll\vF$ (and $a\Omega\ll\vF$), the
kernels varying in $k_0$ on the scale of the matter cone $\vF q$.

\subsection{Magnitudes in the retarded regime}
\label{ssec:numbers}
Let us evaluate the coefficients above in the retarded regime, 
$a\,\omega_0\gg1$: there the round-trip factor confines the Euclidean 
frequencies to
$k_0\lesssim1/(2a)\ll\omega_0$, over which $\alpha(ik_0)\to\alpha(0)$ (this
is what turns the van der Waals $1/a^3$ law into the Casimir-Polder
$1/a^4$ one). For massless graphene the separation $a$ is then the only
scale, and
\begin{equation}
  \delta E^{(0)}=-\,\frac{\alpha(0)}{a^4}\,\mathcal C_0\,,\qquad
  \delta E(\rho;\Omega)-\delta E^{(0)}
  =\frac{\alpha(0)}{a^4}\Big[(\Omega\rho)^2\,\mathcal C_2^{\rm orb}
  +(a\Omega)^2\,\mathcal C_2^{\rm spin}\Big]\,,
  \label{eq:coeffs}
\end{equation}
with velocities in units of $c$; $\delta E^{(0)}$ is the static-sheet shift
\eqref{eq:CP0}, which coincides with $\delta E(\rho;0)$ for every $\rho$
(Sec.~\ref{sec:channels}). Evaluated numerically from \eqref{eq:CP0} and
\eqref{eq:Omega2}, these coefficients are collected in
Table~\ref{tab:graphene}.

\begin{table}[ht]
\centering
\caption{Static and $O(\Omega^2)$ coefficients for a graphene sheet, in
units of $\alpha(0)/a^4$, from Eqs.~\eqref{eq:CP0} and \eqref{eq:Omega2}
with the resummed coefficients \eqref{eq:refl}, for $N=4$ and
$e^2=4\pi/137$ ($\aN\simeq0.0229$), at three Fermi velocities (in units of
$c$).}
\label{tab:graphene}
\begin{tabular}{ccccc}
\toprule
$\vF$ & $\mathcal C_0$ & $\mathcal C_2^{\rm orb}$ & $\mathcal C_2^{\rm spin}$
& $\mathcal C_2^{\rm orb}/\mathcal C_0$ \\
\midrule
$c/100$  & $4.46\times10^{-4}$ & $-2.65\times10^{-4}$ & $-5.43\times10^{-5}$ & $-0.59$\\
$c/300$  & $4.68\times10^{-4}$ & $-2.86\times10^{-4}$ & $-5.75\times10^{-5}$ & $-0.61$\\
$c/1000$ & $4.74\times10^{-4}$ & $-2.91\times10^{-4}$ & $-5.67\times10^{-5}$ & $-0.62$\\
\bottomrule
\end{tabular}
\end{table}

At $\vF=c/300$, $\mathcal C_0$ is about $4.9\%$ of the perfect-mirror value
$3/(32\pi^2)$. The relative $\Omega$ dependence is
\begin{equation}
  \frac{\delta E(\rho;\Omega)-\delta E^{(0)}}{|\delta E^{(0)}|}
  \;\simeq\;-0.61\,\Big(\frac{\Omega\rho}{c}\Big)^{\!2}
  -0.12\,\Big(\frac{a\Omega}{c}\Big)^{\!2}\,.
  \label{eq:relative}
\end{equation}
Since $\delta E^{(0)}<0$, both terms deepen the attractive shift, the
orbital one dominating for $\rho\gtrsim0.45\,a$ and the spin one surviving
alone on the axis. The common sign has a simple Euclidean origin: through
the weight $K^2\partial^2_{k_0}R_l$ both terms probe the same low-frequency
TM shoulder, the concave plateau \eqref{eq:Rl0}, and the
average over the two senses of the shift,
$\half\big[R(k_0+im\Omega)+R(k_0-im\Omega)\big]
=R-\half(m\Omega)^2\partial^2_{k_0}R$, raises a concave function; the
orbital term does so through the large
angular momenta $m\sim q\rho$, the spin term through the single helicity
unit. 

\subsection{Dissipative threshold and induced width}
\label{ssec:threshold}

Beyond the real $O(\Omega^2)$ series the shift develops an imaginary part,
an $\Omega$-induced level width, whose onset follows from the branch cuts.
Continuing
$k_0\to-i\omega$, the rotating coefficients are evaluated at
$\bar\omega=\omega-m\Omega$ and acquire an imaginary part (creation of an
electron-hole pair) only beyond the particle-hole threshold,
$|\bar\omega|>\vF q$; the atomic factor contributes its poles at
$\omega=\pm\omega_0$, with $\omega_0$ the relevant transition frequency.
For the ground-state level to acquire a width, i.e.\ for the static atom to
be spontaneously excited with the rotation as the energy reservoir, the
bookkeeping per exchanged quantum in channel $m$ is: the rotation supplies
$m\Omega$, of which $\omega_0$ excites the atom and the remainder, at least
$\vF q$, creates the pair; the channel must also be available, $|m|\lesssim
q\rho$ at the atom and $q\lesssim1/(2a)$ from the round trip. Combining
these estimates,
\begin{equation}
	\omega_0+\vF q\;<\;m\Omega\,,\qquad |m|\lesssim q\rho\,,\qquad
	q\lesssim\frac{1}{2a}
	\quad\Longrightarrow\quad
	\Omega\rho\;\gtrsim\;\vF+2a\,\omega_0\,.
	\label{eq:threshold}
\end{equation}
Equation \eqref{eq:threshold} should be read as a parametric onset
criterion, the numerical factors being order-one estimates; the strict
reality of $\delta E$ below the onset, however, is exact: the combined
atom-plus-sheet configuration is stationary, the corotating excitation
energies of the modes the atom couples to, those of the patch beneath it,
being positive below \eqref{eq:threshold} by the spectral argument of
Sec.~\ref{sec:sheet}, supplemented by the positive atomic excitation
energies. In the retarded regime $a\omega_0\gg1$ the threshold is
relativistic and the $\Omega$ dependence is purely dispersive. The
interesting regime is the near zone, $a\omega_0\ll\vF$, accessible with
low-frequency transitions (Rydberg pairs, hyperfine or molecular
rotational lines, at sub-micron separations): the threshold then reduces,
parametrically, to the geometric condition that the atom hover beyond the
critical radius
\begin{equation}
	r_c=\frac{\vF}{\Omega}\,,
\end{equation}
where the local speed of the material reaches the Fermi velocity
(Fig.~\ref{fig:geom}). A ground-state atom scanned in $\rho$ across $r_c$
is thus a pointwise probe of the rotating-sheet dissipation: its level is
sharp for $\rho<r_c$ and, beyond it, acquires a width set by the local
response of a sheet sliding at $v=\Omega\rho$, as established for the real
part by the orbital term of Sec.~\ref{sec:smallOmega}. This width is the
spontaneous-excitation rate of atom-surface quantum
friction~\cite{Intravaia}; its explicit form, and its dependence on the
surface loss function and on the chosen material model, are given in
Sec.~\ref{ssec:dissipation}.

\section{Finite-thickness disks: conductors and semiconductors}
\label{sec:drude}
The construction so far has used graphene only as a concrete first example; the
general formula \eqref{eq:masterf} requires as input nothing but the reflection
coefficients of the rotating body. We now apply it to other media,
finite-thickness disks made of an ordinary conductor or of a doped
semiconductor, treating each as an example of the same construction. The
decomposition into angular-momentum channels, the Bessel weights, and the
separation between orbital and spin angular momentum are purely kinematic and
carry over unchanged, provided the atom is not close to the rim. 

What changes is only the material input: the graphene VPT localized at $x_3=0$
must be replaced by the Fresnel reflection amplitudes of a bulk slab. This is
the standard form in which Lifshitz theory, or macroscopic QED near planar
bodies, encodes the material dependence of Casimir-Polder
shifts~\cite{LifshitzPitaevskii,Buhmann,LambrechtScattering,KlimchitskayaRMP}.
For a \emph{bulk} medium the same replacement is a local comoving-frame
(nonrelativistic rotating-scatterer) approximation, the standard prescription of
the non-contact quantum-friction literature~\cite{Volokitin,Intravaia}.
We note, however, that it omits the relativistic response of a moving bulk: the
Roentgen-Fizeau magnetoelectric terms of the Minkowski constitutive relations,
of order $(\epsilon-1)\,\Omega\rho/c$ in amplitude, together with the boost of
the in-plane wavevector and of the polarization basis. By the parity-evenness of
the shift in $\Omega$ these enter only at $O((\Omega\rho/c)^2)$, the nominal
order of the orbital term; but they act through the magnetic field of the
reflected mode and the wavevector boost, both weak for the quasi-electrostatic
near-field modes that dominate the shift. 

We therefore use \eqref{eq:shift} throughout.

\subsection{Planar-slab reduction and Fresnel coefficients}
\label{ssec:slab_coeffs}
Let the disk have radius $\Rdisk$ and thickness $\ell$, with its upper face
at $x_3=0$ and the material occupying $-\ell<x_3<0$. Away from the edge,
$a\ll \Rdisk-\rho$, the field reflected back to the atom probes a region of
lateral size $\sim a$; to that accuracy the disk may be replaced by an
infinite slab of the same thickness. In a more exact finite-radius treatment
one would have to use a cylindrical scattering matrix for a disk. Axial
symmetry would still make the problem diagonal in the conserved angular
momentum $m$, but the radial momentum would no longer be conserved: edge
scattering would mix $q$ and $q'$. The replacement made here is therefore

\begin{equation}
  \hbox{large rotating disk}\quad \longrightarrow \quad
  \hbox{locally planar rotating slab of thickness }\ell,
  \label{eq:disk_to_slab}
\end{equation}
with errors controlled by $a/\Rdisk$ and $a/(\Rdisk-\rho)$, in addition to
possible nonlocal corrections to the dielectric response.

For a local, isotropic dielectric function $\epsilon(\omega)$, define on
the imaginary axis
\begin{equation}
  K=\sqrt{\xi^2+q^2}\,,\qquad
  K_m=\sqrt{\epsilon(i\xi)\,\xi^2+q^2}\,,\qquad \xi>0 .
  \label{eq:bulk_Ks}
\end{equation}
The one-interface amplitudes follow by solving Maxwell's equations in the
two half-spaces and imposing the usual continuity conditions at the planar
interface: $E_\parallel,H_\parallel$ continuous for a dielectric without
free surface charge or current. For TM and TE polarization one obtains,
respectively,
\begin{equation}
  r_p=\frac{\epsilon K-K_m}{\epsilon K+K_m}\,,
  \qquad
  r_s=\frac{K-K_m}{K+K_m}\,.
  \label{eq:bulk_interface}
\end{equation}
These are the Euclidean Fresnel coefficients. They should not be confused
with the positive quantities $R_l,R_t$ used earlier for graphene: in the
present sign convention
\begin{equation}
  R_l=r_p\,,\qquad R_t=-r_s
  \label{eq:bulk_sign_convention}
\end{equation}
for a single interface. The minus sign in the TE channel is the same one
that led to $r_{\rm te}=-R_t$ in Eq.~\eqref{eq:refl}; it is a convention
chosen so that the electric-field trace in Eq.~\eqref{eq:CP0} contains
$+k_0^2R_t$.

The finite thickness is included by summing the Fabry-Perot series of
multiple internal reflections in the slab. The incident wave reflects at the
upper face with amplitude $r_\sigma$, while the part transmitted into the
medium crosses the slab, partially reflects at the lower face, and returns to
escape from above, repeatedly. Two ingredients control the sum: by the Stokes
relations the reflection seen from inside the medium is $-r_\sigma$ at either
face and the transmissions obey $t_\sigma t_\sigma'=1-r_\sigma^2$; and, on the
imaginary-frequency axis, a single crossing of the slab multiplies the
amplitude by the real decay factor $e^{-K_m\ell}$ (replacing the oscillatory
phase of real frequencies), so a full internal round trip costs
$e^{-2K_m\ell}$. Summing the direct reflection and the geometric series of
escaping internal paths, $r_\sigma+(1-r_\sigma^2)(-r_\sigma)e^{-2K_m\ell}
\sum_{n\ge0}(r_\sigma^2 e^{-2K_m\ell})^n$, gives, since the material is vacuum
on both sides, the reflection coefficient seen from the upper vacuum region,
\begin{equation}
  r_\sigma^{(\ell)}
  =\frac{r_\sigma\big(1-e^{-2K_m\ell}\big)}
         {1-r_\sigma^2 e^{-2K_m\ell}}\,,
  \qquad \sigma=p,s .
  \label{eq:slab_FabryPerot}
\end{equation}
Thus, in the notation of the rest of the paper,
\begin{equation}
  R_l(\xi,q;\ell)
  =\frac{r_p\big(1-e^{-2K_m\ell}\big)}
         {1-r_p^2 e^{-2K_m\ell}}\,,
  \qquad
  R_t(\xi,q;\ell)
  =-\frac{r_s\big(1-e^{-2K_m\ell}\big)}
          {1-r_s^2 e^{-2K_m\ell}}\,.
  \label{eq:bulk_slab}
\end{equation}
Equations~\eqref{eq:bulk_interface} and \eqref{eq:bulk_slab} are the usual
thin-film/slab replacement used in Lifshitz calculations with finite
thickness~\cite{PirozhenkoLambrecht,KlimchitskayaRMP}. The limit
$\ell\to\infty$ gives a half-space. The opposite regime, $q\ell\ll1$,
would reduce the body to an effective conducting film; that is a different
model from the thick-disk limit meant here.

The rotating-body formula has the same angular structure as
Eq.~\eqref{eq:masterf}:
\begin{align}
  \delta E_{\rm bulk}(\rho;\Omega)=&-\sum_{m=-\infty}^{\infty}
  \int_{-\infty}^{\infty}\frac{dk_0}{2\pi}
  \int_0^\infty\frac{q\,dq}{2\pi}\,\alpha(ik_0)
  \frac{e^{-2aK}}{4K}
  \Big\{q^2J_m^2(q\rho)\,R_l(\bar k_0,q;\ell)
  \nonumber\\
  &+\frac{1}{2}\big[J_{m-1}^2(q\rho)+J_{m+1}^2(q\rho)\big]
  \times\big[K^2R_l+k_0^2R_t\big](\bar k_0,q;\ell)
  \Big\}\,,
  \label{eq:BulkMaster}
\end{align}
where, as before, $\bar k_0=k_0+im\Omega$ and the vacuum propagation factor
$K=\sqrt{k_0^2+q^2}$ is a laboratory quantity. In Eq.~\eqref{eq:BulkMaster}
the shifted argument is understood through the analytic continuation of
the retarded reflection amplitudes; for real Euclidean $k_0$ it reduces to
Eqs.~\eqref{eq:bulk_interface}-\eqref{eq:bulk_slab}. The point of writing
the result in this form is that all operations are now explicit: choose a
causal bulk dielectric function $\epsilon(\omega)$, compute the
one-interface Fresnel amplitudes, dress them by the slab Fabry-Perot factor,
and finally apply the same angular Doppler shift as for graphene.

\subsection{Drude disk}
\label{ssec:drude_disk}

For a local Drude metal~\cite{KlimchitskayaRMP,LambrechtScattering}
\begin{equation}
  \epsilon_{\rm D}(\omega)
  =1-\frac{\omega_p^2}{\omega(\omega+i\gamma)}\,,
  \qquad
  \epsilon_{\rm D}(i\xi)
  =1+\frac{\omega_p^2}{\xi(\xi+\gamma)}\,,
  \qquad \xi>0,
  \label{eq:drudeeps}
\end{equation}
where $\omega_p$ is the plasma frequency and $\gamma$ the relaxation rate.
The coefficients $R_{l,t}^{\rm D}$ are obtained by inserting
$\epsilon=\epsilon_{\rm D}$ in Eqs.~\eqref{eq:bulk_interface} and
\eqref{eq:bulk_slab}.

Several qualitative changes follow. First, consider the small-$\Omega$
expansion. The orbital and spin decomposition of \eqref{eq:Omega2} rests
only on the Bessel sum rules \eqref{eq:sumrules}, which are kinematic, so
it carries over verbatim with $R_{l,t}\to R_{l,t}^{\rm D}$ provided the
$O(\Omega^2)$ coefficients exist. 
A point to keep in view is that the relevant derivative is that of the causal
analytic continuation, not of the real-axis $|k_0|$ expansion. 
The Doppler shift  continues $k_0\to k_0+im\Omega$ off
the real axis, where $\epsilon_{\rm D}(ik_0)=1+\omega_p^2/[k_0(k_0+\gamma)]$
is meromorphic with simple poles at $k_0=0,-\gamma$ away from the contour, so
$R_l^{\rm D}\to1$ \emph{smoothly} along the shifted contour. The correct
$\partial_{k_0}^2R_s^{\rm D}$ is the second derivative of that continuation.
The expansion is no longer universal: the coefficients
are functions of $a\omega_p$ and $a\gamma$.

In practice the coefficients enter \eqref{eq:Omega2} through three explicit
steps. (i) From the retarded permittivity \eqref{eq:drudeeps}, build the
reflection coefficients as analytic functions of a complex frequency
argument $z$, keeping the denominator in its smooth rational form,
\begin{equation}
  \epsilon_{\rm D}(iz)=1+\frac{\omega_p^2}{z(z+\gamma)}\,,\qquad
  K_m=\sqrt{\epsilon_{\rm D}(iz)\,z^2+q^2}\,,
  \label{eq:epscont}
\end{equation}
with $R_l^{\rm D}(z,q),R_t^{\rm D}(z,q)$ from
\eqref{eq:bulk_interface}-\eqref{eq:bulk_slab}; on the positive real axis
$z=k_0>0$ these reduce to the physical imaginary-axis coefficients. (ii) The
weights in \eqref{eq:Omega2} are the second derivatives with respect to the
frequency argument, evaluated on that axis,
\begin{equation}
  \partial_{k_0}^2R_s^{\rm D}(k_0,q)
  =\frac{d^2}{dz^2}R_s^{\rm D}(z,q)\Big|_{z=k_0}\,,\qquad k_0>0\,.
  \label{eq:d2recipe}
\end{equation}
This is the analytic continuation of the response off the imaginary axis,
the angular Doppler shift evaluating $R_s^{\rm D}$ at the complex
\emph{physical} frequency $\bar\omega=-m\Omega+ik_0$. Numerically the same 
object follows from an imaginary displacement,
\begin{equation}
  \partial_{k_0}^2R_s^{\rm D}
  =\frac{R_s^{\rm D}(k_0+i\delta,q)+R_s^{\rm D}(k_0-i\delta,q)
        -2R_s^{\rm D}(k_0,q)}{(i\delta)^2}+O(\delta^2)\,,
  \label{eq:d2num}
\end{equation}
the two sample points lying at $\mathrm{Re}\,\omega=\mp\delta$,
$\mathrm{Im}\,\omega=k_0>0$, in the upper half-plane where
$\epsilon_{\rm D}$ is analytic; note the denominator $(i\delta)^2=-\delta^2$.
(iii) Insert into \eqref{eq:Omega2} and fold the frequency integral onto the
positive axis,
\begin{equation}
  \int_{-\infty}^{\infty}\!dk_0\,(\cdots)=2\int_0^\infty\!dk_0\,(\cdots)\,,
  \label{eq:fold}
\end{equation}
the integrand being even by the reality $\epsilon(-i k_0)=\epsilon(ik_0)$ on
the axis. That a moving medium must be treated through this
continuation, with its altered analytic structure in the complex-frequency
plane, is the resolution of the Philbin-Leonhardt controversy over quantum
friction~\cite{PhilbinLeonhardt,VolokitinComment}; the spectroscopic version
with complex Doppler-shifted frequencies is that of Ref.~\cite{KlattBuhmann}.

Evaluating \eqref{eq:Omega2} with the continued Drude coefficients, 
for an off-axis atom, gives at $a=100$~nm
\begin{equation}
  \frac{\delta E_{\rm D}(\rho;\Omega)-\delta E_{\rm D}^{(0)}}
       {|\delta E_{\rm D}^{(0)}|}
  \simeq -0.16\,\Big(\frac{\Omega\rho}{c}\Big)^{\!2}
  -0.06\,\Big(\frac{a\Omega}{c}\Big)^{\!2}
  \qquad(\text{gold; copper within }2\%)\,,
  \label{eq:DrudeRelative}
\end{equation}
to be compared with $-0.61(\Omega\rho/c)^2-0.12(a\Omega/c)^2$ for graphene,
Eq.~\eqref{eq:relative}. The two have the \emph{same} sign: as for graphene,
the rotation deepens the attraction. This is the expected behavior, since
both responses share a concave low-frequency TM shoulder, although of
different height: a conductor reaches $R_l\to1$ as $k_0\to0$ (a perfect
electrostatic reflector, $\epsilon\to\infty$), whereas graphene saturates at
the subunity plateau \eqref{eq:Rl0}, its Dirac sea screening static in-plane
fields only partially. What the two share, and what fixes the common sign, is
the concave curvature of that shoulder, raised by the Euclidean Doppler
average (Sec.~\ref{ssec:numbers}). The contrast is
one of magnitude rather than sign: the metallic \emph{relative} effect is
about four times smaller than graphene's, not because the absolute
rotational coefficients are smaller (they are in fact several times larger)
but because the static shift $\mathcal C_0$ against which they are measured
is some fifteen times bigger. (The values \eqref{eq:DrudeRelative} are
computed from the half-space coefficients; finite thickness multiplies both
by an $\ell$-dependent factor that tends to one as $\ell\to\infty$.) Should
$\gamma$ be small enough that the cusp is sampled only at frequencies below
the round-trip cutoff $1/2a$, i.e.\ $a\omega_p^2/\gamma\gg1$, the
neighborhood of $k_0=0$ dominates and the unexpanded master formula
\eqref{eq:BulkMaster}, or an equivalent real-frequency form with the
retarded permittivity, is the safer numerical starting point; the
coefficients above already use it.

Second, consider the overall magnitude, set by the static coefficient. In
the retarded regime,
\begin{equation}
  \delta E_{\rm D}^{(0)}=-\frac{\alpha(0)}{a^4}\,
  \mathcal C_0^{\rm D}(a\omega_p,a\gamma,\ell/a)\,,
  \label{eq:DrudeC0}
\end{equation}
and similarly for the rotational correction. The cleanest benchmark, which
removes the disk radius, the thickness and the detailed atomic spectrum, is
the half-space static coefficient $\mathcal C_0^{\rm D}(a\omega_p,a\gamma)$,
directly comparable with the graphene number $\mathcal C_0\simeq
4.7\times10^{-4}$ of Sec.~\ref{ssec:numbers}.

For a good conductor the TM coefficient approaches the perfect-reflector
value over most of the sampled frequency range, while the TE
zero-frequency contribution is suppressed by the Drude prescription
($R_t^{\rm D}\to0$ as $\xi\to0$). The static coefficient is then a sizeable
fraction of the perfect-mirror ceiling $\mathcal C_0^{\rm
mirror}=3/(32\pi^2)\simeq9.5\times10^{-3}$, in sharp contrast to graphene,
where $\mathcal C_0$ is only $4.9\%$ of that ceiling, the smallness being
set by the coupling $\alpha_N\simeq0.023$. Using standard Casimir-model
parameters, the half-space Drude coefficient $\mathcal C_0^{\rm D}$
evaluates to the values collected in Table~\ref{tab:GoldC0}.

\begin{table}[ht]
\centering
\caption{Half-space static Drude coefficient $\mathcal C_0^{\rm D}$ for gold
and copper, in units of $\alpha(0)/a^4$, at three atom-surface separations,
obtained from the imaginary-axis Fresnel coefficients
\eqref{eq:bulk_interface} with $\epsilon=\epsilon_{\rm D}$.}
\label{tab:GoldC0}
\begin{tabular}{lccc}
\toprule
 & $a=50$~nm & $a=100$~nm & $a=200$~nm\\
\midrule
gold ($\omega_p=9.0$~eV, $\gamma=0.035$~eV)
 & $5.8\times10^{-3}$ & $7.1\times10^{-3}$ & $8.1\times10^{-3}$\\
copper ($\omega_p=8.8$~eV, $\gamma=0.030$~eV)
 & $5.7\times10^{-3}$ & $7.1\times10^{-3}$ & $8.1\times10^{-3}$\\
\bottomrule
\end{tabular}
\end{table}

With $a\omega_p\simeq2.3$, $4.6$ and $9.1$ for gold (and almost the same for
copper), these are $61\%$, $75\%$ and $85\%$ of the perfect-mirror value
respectively, and roughly $12$ to $17$ times the graphene coefficient. The
two metals differ by less than a percent at these separations; what matters
is that the metallic shift approaches the ideal-conductor limit as
$a\omega_p$ grows, whereas the graphene shift is permanently held down by
$\alpha_N$. The plasma model gives values within a fraction of a percent of
the Drude ones for $\mathcal C_0$ (the static coefficient being insensitive
to the $\xi\to0$ behavior that distinguishes the two models), so the same
numbers serve as the good-conductor benchmark for both.

Finally, the rotational correction. Because the continued Drude
coefficients are analytic along the shifted contour, the $O(\Omega^2)$
formula \eqref{eq:Omega2} applies as discussed above, both for a half-space
and for a slab of finite thickness $\ell$ through the Fabry-Perot factor
\eqref{eq:bulk_slab}. At $a=100$~nm (gold), in units of $\alpha(0)/a^4$,
\begin{equation}
  \mathcal C_2^{\rm orb,D}\simeq-1.13\times10^{-3}\,,\qquad
  \mathcal C_2^{\rm spin,D}\simeq-4.24\times10^{-4}
  \label{eq:DrudeC2}
\end{equation}
for the half-space (copper within $2\%$); finite thickness leaves these
essentially unchanged until $\ell$ drops below a few skin depths
$\omega_p^{-1}$ ($\simeq0.2\,a$ here), below which all coefficients are
suppressed together. Both coefficients share the sign of the graphene ones,
as displayed by the relative shift \eqref{eq:DrudeRelative}: the rotation
deepens the attraction to a Drude conductor as well. We stop the model at
this order. Below it the shift is real and given by \eqref{eq:Omega2}; an
imaginary part, signaling spontaneous excitation, requires the material to
absorb at the Doppler-shifted frequency, and since
$\mathrm{Im}\,\epsilon_{\rm D}(\omega)>0$ for all $\omega>0$ a Drude metal
has no protective gap analogous to the graphene matter cone, so that
dissipative regime lies outside the $O(\Omega^2)$ description and is not
pursued here.

\subsection{Plasma disk}
\label{ssec:plasma_disk}

The plasma model, often used as the dissipationless alternative in Casimir calculations~\cite{KlimchitskayaRMP,HartmannIngold}, is obtained by suppressing Ohmic relaxation from the
outset,
\begin{equation}
  \epsilon_{\rm P}(\omega)=1-\frac{\omega_p^2}{\omega^2}\,,
  \qquad
  \epsilon_{\rm P}(i\xi)=1+\frac{\omega_p^2}{\xi^2}\,,
  \qquad \xi>0 .
  \label{eq:plasmaeps}
\end{equation}
Equivalently,
\begin{equation}
  K_{\rm P}=\sqrt{q^2+\xi^2+\omega_p^2}\,.
\end{equation}
The coefficients $R_{l,t}^{\rm P}$ are again given by
Eqs.~\eqref{eq:bulk_interface}-\eqref{eq:bulk_slab}, now with
$\epsilon=\epsilon_{\rm P}$ and $K_m=K_{\rm P}$. The master formula is
therefore Eq.~\eqref{eq:BulkMaster} with
$R_{l,t}\to R_{l,t}^{\rm P}$.

This is not identical to taking the Drude result and then sending
$\gamma\to0$. The order of the limits $\xi\to0$ and $\gamma\to0$ matters.
For a plasma half-space,
\begin{equation}
  R_l^{\rm P}(\xi,q)
  =1-\frac{2\sqrt{q^2+\omega_p^2}}{q\omega_p^2}\,\xi^2+O(\xi^4)\,,
  \qquad
  R_t^{\rm P}(0,q)
  =\frac{\sqrt{q^2+\omega_p^2}-q}
         {\sqrt{q^2+\omega_p^2}+q}\,.
  \label{eq:PlasmaLowFreq}
\end{equation}
Thus the TM coefficient again reaches the perfect-conductor value at zero
frequency, but now in an analytic, even way; and, unlike the Drude model,
the TE coefficient does not vanish at zero frequency. This is the usual
Drude-plasma discontinuity of the local description, here appearing in
the rotational correction as well.

Consequently the small-$\Omega$ expansion is regular in the plasma model,
with no subtlety of analytic continuation: $\epsilon_{\rm P}(ik_0)
=1+\omega_p^2/k_0^2$ is even and analytic on the real Euclidean axis away
from $k_0=0$, and $R_l^{\rm P}$ approaches its zero-frequency plateau
quadratically rather than through a cusp. Equation~\eqref{eq:Omega2}
carries over directly with $R_{l,t}\to R_{l,t}^{\rm P}$, again splitting
into an orbital term $\propto(\Omega\rho)^2$ and a spin term
$\propto(a\Omega)^2$, both for a half-space and for a slab of finite
thickness $\ell$. The static coefficient and, with it, the rotational ones
are no longer pure numbers but functions of $a\omega_p$ (and $\ell/a$):
\begin{equation}
  \delta E_{\rm P}^{(0)}=-\frac{\alpha(0)}{a^4}\,
  \mathcal C_0^{\rm P}(a\omega_p,\ell/a)\,,
\end{equation}
and likewise for $\mathcal C_2^{\rm orb,P}$, $\mathcal C_2^{\rm spin,P}$.
The static coefficient is insensitive to the $k_0\to0$ behavior that
distinguishes the two models, so $\mathcal C_0^{\rm P}$ agrees with the
Drude values of Table~\ref{tab:GoldC0} to within a fraction of a percent. The
rotational coefficients are likewise close to the Drude ones: at
$a=100$~nm (gold), in units of $\alpha(0)/a^4$,
\begin{equation}
  \mathcal C_2^{\rm orb,P}\simeq-1.12\times10^{-3}\,,\qquad
  \mathcal C_2^{\rm spin,P}\simeq-4.00\times10^{-4}
  \label{eq:PlasmaC2}
\end{equation}
for the half-space, so that the relative shift is again
\begin{equation}
  \frac{\delta E_{\rm P}(\rho;\Omega)-\delta E_{\rm P}^{(0)}}
       {|\delta E_{\rm P}^{(0)}|}
  \simeq -0.16\,\Big(\frac{\Omega\rho}{c}\Big)^{\!2}
  -0.06\,\Big(\frac{a\Omega}{c}\Big)^{\!2}\,,
  \label{eq:PlasmaRelative}
\end{equation}
of the same sign as Drude and as graphene. In the good-conductor
limit $a\omega_p\gg1$ both polarizations approach the perfect-reflector
result, finite plasma frequency producing corrections controlled by the
skin depth $\omega_p^{-1}$ and finite thickness by $\ell/a$. We stop the
plasma model here as well: being lossless, it has $\mathrm{Im}\,
\epsilon_{\rm P}=0$, so the $O(\Omega^2)$ shift \eqref{eq:PlasmaRelative}
is real and a width can arise only non-perturbatively, when the rotation
feeds a real photonic or surface-plasmon mode of the slab, again outside
the present order.

In short, replacing graphene by a thin or thick conducting plate leaves the
angular-momentum architecture of the calculation, and the orbital-plus-spin
structure of the $O(\Omega^2)$ shift, intact; what changes is the material
input. The graphene coefficients are pure numbers, fixed by $\alpha_N$ and
$\vF$, and small ($\mathcal C_0$ is $4.9\%$ of the perfect-mirror value);
the metallic coefficients depend on $a\omega_p$ (and $a\gamma$, $\ell/a$)
and approach the ideal-conductor limit as $a\omega_p$ grows. In all three
cases the rotation deepens the attraction: the metallic and graphene
coefficients share their sign, fixed by the common concave low-frequency TM
shoulder, and the two metallic models give nearly identical
$O(\Omega^2)$ coefficients because they differ only in the immediate
neighborhood of $k_0=0$, which contributes little to the shift. What
distinguishes graphene is not the sign but the magnitude: its anomalously
small static shift makes the relative rotational effect several times larger
than for a good conductor.

\subsection{Doped semiconductors: reversal of the sign}
\label{ssec:semiconductor}

The three media examined so far, graphene and the Drude and plasma metals, all
\emph{deepen} the attraction under rotation, with a common origin: the
concave low-frequency TM shoulder that controls the near-field band
$k_0\sim1/a$. This behavior is not universal. A doped semiconductor carries a
free-carrier (Drude) response superposed on the lattice background
$\epsilon_\infty$,
\begin{equation}
  \epsilon(ik_0)=\epsilon_\infty+\frac{\omega_p^2}{k_0(k_0+\gamma)},
  \qquad \omega_p^2=\frac{n e^2}{\epsilon_0 m^*},
  \label{eq:epsSi}
\end{equation}
with $\omega_p$ the unscreened carrier plasma frequency set by the doping
density $n$ and the conductivity effective mass $m^*$. For n-type silicon
($\epsilon_\infty\simeq11.7$, $m^*\simeq0.26\,m_e$) at a heavy doping
$n=10^{20}\,\mathrm{cm^{-3}}$ and mobility $\mu\simeq70\,\mathrm{cm^2/(V\, s)}$,
values representative of heavily doped n-Si and consistent with the
experimental proposal of Ref.~\cite{DecaLombardo}, one
has $\omega_p\simeq0.73$~eV and $\gamma\simeq0.06$~eV, that is
$a\omega_p\simeq0.37$ and $a\gamma\simeq0.03$ at $a=100$~nm. Evaluating the
coefficients of Appendix~\ref{app:bulkderiv} for this $\epsilon(ik_0)$ gives
\begin{equation}
  \mathcal C_0\simeq6.6\times10^{-3},\qquad
  \mathcal C_2^{\rm orb}\simeq+1.6\times10^{-3},\qquad
  \mathcal C_2^{\rm spin}\simeq+3.8\times10^{-4},
  \label{eq:SiC2}
\end{equation}
so that $\mathcal C_2^{\rm orb}/\mathcal C_0\simeq+0.25$. Both $O(\Omega^2)$
coefficients are now \emph{positive}: the rotation \emph{weakens} the
attraction, opposite to the metals and to graphene.

\begin{figure}[t]
  \centering
  \includegraphics[width=0.72\textwidth]{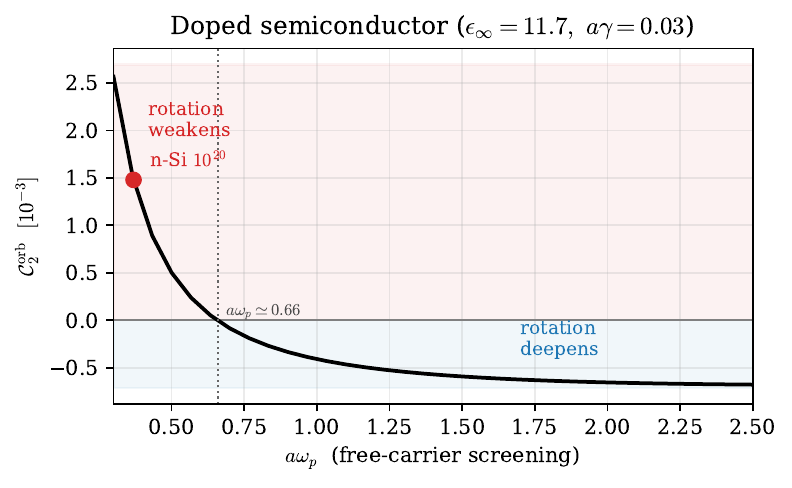}
  \caption{Orbital coefficient $\mathcal C_2^{\rm orb}$ for the
  doped-semiconductor model \eqref{eq:epsSi} versus the free-carrier screening
  parameter $a\omega_p$, at fixed $\epsilon_\infty=11.7$ and $a\gamma=0.03$. The
  coefficient is positive (rotation weakens the attraction) when the carrier
  plasma frequency lies below the near-field scale, $a\omega_p\lesssim0.66$, and
  turns negative (metallic behavior, rotation deepens the attraction) above it.
  The marker is heavy n-type silicon, $n=10^{20}\,\mathrm{cm^{-3}}$ at
  $a=100$~nm.}
  \label{fig:semicrossover}
\end{figure}

The sign is governed by the curvature of $R_l$ across the near-field band. A
good conductor has $a\omega_p\gg1$, so the shoulder $R_l\to1$ spans the band and
the response rolls off concavely, $\partial_{k_0}^2R_l<0$, giving
$\mathcal C_2<0$. In n-Si the carrier plasma frequency lies \emph{below} the
band, $a\omega_p<1$: the screening shoulder is confined to
$k_0\lesssim\omega_p$, and within the band the response samples the
\emph{convex} rise toward that shoulder, reversing the curvature and hence the
sign. The discriminator is simply whether the carrier screening reaches into the
near-field band, $\omega_p\gtrless1/a$. Figure~\ref{fig:semicrossover} shows
$\mathcal C_2^{\rm orb}$ for the model \eqref{eq:epsSi} as the carrier density is
varied at fixed $\epsilon_\infty$ and $\gamma$: it is positive for
$a\omega_p\lesssim0.66$ (under-screened, attraction weakened) and crosses to the
metallic branch $\mathcal C_2<0$ for $a\omega_p\gtrsim0.66$, i.e.
$n\gtrsim3\times10^{20}\,\mathrm{cm^{-3}}$ at $a=100$~nm. The pure lattice
dielectric ($\omega_p\to0$, $R_l\to(\epsilon_\infty-1)/(\epsilon_\infty+1)<1$)
has no shoulder and returns a small negative coefficient; the positive sign is
therefore specific to a partially screened conductor, where the shoulder exists
but sits below the band.

The practical implication is that the sign of the rotational Lamb shift tracks
the metal-insulator crossover of the substrate. Tuning the doping through
$\omega_p\sim1/a$, or equivalently the separation through $a\sim1/\omega_p$,
weakens the rotation-induced shift, which turns from binding to anti-binding, a 
control knob absent in the fixed-sign metallic and graphene cases.

\subsection{Dissipation: Drude versus plasma}
\label{ssec:dissipation}

The near-coincidence of the real $O(\Omega^2)$ coefficients does not make the
two metal models physically equivalent: they differ qualitatively in their
\emph{dissipative} behavior, i.e.\ in the imaginary part of the shift. A
nonzero $\mathrm{Im}\,\delta E$ endows the ground state with a finite
excitation rate $\Gamma=-2\,\mathrm{Im}\,\delta E$, the rotation supplying the
energy; it is the spectroscopic image of quantum
friction~\cite{KlattBuhmann,Intravaia}, here with the surface, rather than the
atom, in motion. (A ground-state atom facing a \emph{static} passive surface
has $\mathrm{Im}\,\delta E=0$ at $T=0$: a lossy surface broadens only the
excited levels.)

The imaginary part is controlled by the surface loss function
$\mathrm{Im}\,R_s(\bar\omega,q)$ at the Doppler-shifted real frequency
$\bar\omega=\omega-m\Omega$, and here the two models part ways. The Drude
response is Ohmic and gapless,
\begin{equation}
  \mathrm{Im}\,\epsilon_{\rm D}(\omega)
  =\frac{\omega_p^2\,\gamma}{\omega(\omega^2+\gamma^2)}>0
  \qquad(\omega>0),
  \label{eq:ImepsD}
\end{equation}
so $\mathrm{Im}\,R_l^{\rm D}\neq0$ at \emph{every} real frequency, with no
matter-cone gap; the plasma permittivity is real, $\mathrm{Im}\,\epsilon_{\rm
P}=0$, so $\mathrm{Im}\,R_s^{\rm P}=0$ at every real frequency off its
surface-mode resonances and the plasma plate carries no \emph{Ohmic} width.
This last statement needs a caveat. The plasma reflection coefficient
$R_l^{\rm P}$ has surface-plasmon-polariton poles at real frequencies
$\omega_{\rm sp}(q)$; should the Doppler-shifted frequency
$\bar\omega=\omega_0-m\Omega$ be brought onto one of them, the rotation can
resonantly excite a surface plasmon and open a width even with
$\mathrm{Im}\,\epsilon_{\rm P}=0$.
\section{Conclusions}
\label{sec:concl}

We have shown that the $\Omega$-dependent Lamb shift of a static atom
facing a rotating planar surface is governed by a single object: the
angularly Doppler-shifted reflection coefficients of the surface, probed
pointwise by the atom. The general formula for the shift,
Eq.~\eqref{eq:masterf}, is diagonal in the total angular momentum of the
photon modes, with Bessel weights giving the probability that a photon of
given in-plane momentum, observed at lateral distance $\rho$ from the axis,
carries angular momentum $m$.

From that formula we obtained a closed $O(\Omega^2)$ result,
Eq.~\eqref{eq:Omega2}, which splits into a local-density orbital term
$\propto(\Omega\rho)^2$, the sliding result at the local velocity, and a
polarization term $\propto(a\Omega)^2$, the rotational Doppler shift of
the photon helicity; for a point probe, the two terms realize the leading
order and the first corrections of the derivative expansion in the
velocity field. 

The spatial profile of the rotational Lamb shift has interesting consequences: 
the $a$-dependence introduces $\Omega$-dependent corrections to the standard 
Casimir-Polder attraction normal to the surface. More notably, the $\rho$-dependence (driven 
by the orbital angular momentum of the exchanged photons) induces a radial force. This lateral force, which is independent of the sense of rotation of the disk, is a distinctive feature of the rotating system.

We have also shown that the shift is strictly real below a threshold, the
level acquiring an $\Omega$-induced width once the rotation can supply, in
some channel, the atomic excitation energy together with any matter-cone
cost of pair creation; for graphene this is
$\Omega\rho\gtrsim\vF+2a\omega_0$, and for low-frequency transitions it
makes the atom a local probe of the annulus $\rho>r_c=\vF/\Omega$, where
the rotating surface behaves locally as a dissipative, sliding medium. The friction force is in the azimuthal direction, and its sign depends on the sense of rotation of the disk.

We introduced the method on graphene and then applied the same formula to
finite-thickness conducting and semiconducting disks, each an example of
the construction with its own reflection coefficient. The angular-momentum
machinery and the orbital-plus-spin decomposition of the $O(\Omega^2)$
shift are unchanged; only the material response differs, that of a bulk
slab (one-interface Fresnel amplitudes dressed by a Fabry-Perot factor)
replacing the two-dimensional graphene VPT. For graphene and for the Drude
and plasma metals the rotation deepens the Casimir-Polder attraction, the
sign being fixed by the common concave low-frequency TM shoulder (which for
the metals reaches unity, and for graphene saturates below it); the
metallic coefficients, no longer pure numbers, depend on $a\omega_p$ (and
$a\gamma$, $\ell/a$) and approach the perfect-reflector limit as
$a\omega_p$ grows. A doped semiconductor breaks this pattern: when the
carrier plasma frequency lies below the near-field scale, $a\omega_p<1$, as
for heavily doped n-type silicon, the sign reverses and the rotation
\emph{weakens} the attraction. The sign of the rotational shift thus tracks
the metal-insulator crossover of the substrate, a tunable feature with no
analog in the fixed-sign metallic and graphene cases. Dissipation provides
a further discriminant: only the Drude plate gives the rotating atom a
finite width through its Ohmic loss continuum, the lossless plasma having
none, a resonant surface-mode channel aside.

Natural extensions of this work include finite temperature, as well as
gapped or magnetically biased sheets, where a Hall term yields
linear-in-$\Omega$ effects already for isotropic atoms.

\section*{Acknowledgments}
This work was supported by Consejo Nacional de Investigaciones
Científicas y Técnicas (CONICET), Universidad Nacional de Cuyo (UNCuyo), and Universidad de
Buenos Aires (UBA).

\appendix
\section{Solution of the Maxwell equations for a point dipole facing the
rotating sheet}
\label{app:scattering}

In this Appendix we derive the general formula \eqref{eq:masterf} by solving
explicitly the classical problem behind the reflection picture of
Secs.~\ref{sec:static} and \ref{sec:channels}: the Maxwell equations for the
field radiated by a point dipole located at $\rA$, in the presence of the
rotating sheet. The field reflected by the sheet back to the position of the
dipole determines, by linear response, the scattering part of the correlator
that enters Eq.~\eqref{eq:WS}, and Eq.~\eqref{eq:masterf} follows. The
procedure is the one of Sec.~\ref{sec:static}, where the static sheet was
treated in plane waves, now carried out in the cylindrical modes of
Sec.~\ref{sec:channels}, which are the ones the rotating sheet reflects
without mixing. Solutions of the Maxwell equations in the presence of
rotating bodies have a long history, from the low-velocity electrodynamics of
rotating conductors and dielectrics \cite{VanBladel,Shiozawa} to the
reflection of cylindrical waves by a rotating cylinder, where the shift
$\omega\to\omega-m\Omega$ of the angular-momentum channels was first
identified \cite{Zeldovich,BekensteinSchiffer}, the explicit scattering
solutions for rotating spheres and cylinders \cite{DeZutter}, and the
scattering-theory formulation of the quantum radiation of rotating objects
\cite{MaghrebiRot}; see also the experiments of Ref.~\cite{RotExp}. The
treatment below is the planar version of those cylindrical-wave analyses,
with the response of the body entering through its surface current.

\subsection{Field equations and matching conditions}
\label{app:matching}

We work at real frequency, with time dependence $e^{-i\omega t}$, in the
Heaviside-Lorentz units of the text ($c=1$). A point electric dipole $\mathbf
d$ at $\rA=(\boldsymbol\rho_A,a)$, $a>0$, is the polarization source $\mathbf
P=\mathbf d\,\delta^3(\mathbf r-\rA)$, and the Maxwell equations away from the
sheet give
\begin{equation}
  \nabla\times\nabla\times\mathbf E-\omega^2\mathbf E
  =\omega^2\,\mathbf d\,\delta^3(\mathbf r-\rA)\,,\qquad
  \mathbf B=\frac{1}{i\omega}\,\nabla\times\mathbf E\,,\qquad x_3\neq0\,.
  \label{eq:appMaxwell}
\end{equation}
The sheet occupies the plane $x_3=0$ and carries the surface current $\mathbf
J_s$ induced in it by the field. Integrating the Faraday and Amp\`ere laws
across the plane gives the matching conditions
\begin{equation}
  \hat z\times\big[\mathbf E\big]_{0^-}^{0^+}=0\,,\qquad
  \hat z\times\big[\mathbf B\big]_{0^-}^{0^+}=\mathbf J_s\,,
  \label{eq:appmatch}
\end{equation}
$[\,\cdot\,]_{0^-}^{0^+}$ denoting the jump across the sheet. The solution is
the sum of the field the dipole radiates in free space and of the field
reflected by the sheet, $\mathbf E=\mathbf E^{(0)}+\mathbf E^{\rm sc}$, the
latter regular at $\mathbf r=\rA$ and linear in $\mathbf d$.

The problem is closed by the constitutive relation of the sheet, which relates
$\mathbf J_s$ to the field on the plane. For the sheet at rest this is the
linear response of Sec.~\ref{sec:sheet}, written in terms of the tangential
electric field. In the plane-wave basis it reads
\begin{equation}
  \tilde{\mathbf J}_s(\omega,\kq)
  =\big[\sigma_l(\omega,q)\,\hat k\hat k+\sigma_t(\omega,q)\,\hat e\hat e\big]
  \cdot\tilde{\mathbf E}_\parallel(\omega,\kq;x_3=0)\,,
  \qquad \hat e=\hat z\times\hat k\,,
  \label{eq:appsigma}
\end{equation}
so that $\sigma_l$ acts on the irrotational part of the in-plane field and
$\sigma_t$ on its solenoidal part; the two projectors are those of
Eq.~\eqref{eq:VPTdecomp}, restricted to the tangential electric field. The
conductivities and the VPT kernels describe the same response and are
related, on the imaginary axis, by
\begin{equation}
  \sigma_t(ik_0,q)=\frac{g_t(k_0,q)}{k_0}\,,\qquad
  \sigma_l(ik_0,q)=\frac{k_0\,g_l(k_0,q)}{K^2}\,,
  \label{eq:appdict}
\end{equation}
as we verify below by recovering Eq.~\eqref{eq:refl}. For graphene,
Eqs.~\eqref{eq:gt}-\eqref{eq:gl} give
$\sigma_t=\aN\sqrt{k_0^2+\vF^2q^2}/k_0$ and
$\sigma_l=\aN k_0/\sqrt{k_0^2+\vF^2q^2}$, both reducing to the universal
conductivity $\aN$ (in these units) in the local limit $q\to0$.

For the rotating sheet the constitutive relation is that of the material at
rest in the comoving frame, as stated in Sec.~\ref{sec:sheet}. We make
explicit how it acts on a laboratory field of definite frequency and total
angular momentum. Let $\mathcal R(\beta)$ denote the rotation by $\beta$ about
the axis. A field $\mathbf E(t,\boldsymbol\rho)=e^{-i\omega t}\,\mathbf
e_m(\boldsymbol\rho)$ on the plane has total angular momentum $m$ if
\begin{equation}
  \mathcal R(-\beta)\,\mathbf e_m\big(\mathcal R(\beta)\boldsymbol\rho\big)
  =e^{im\beta}\,\mathbf e_m(\boldsymbol\rho)\,,
  \label{eq:appJz}
\end{equation}
the rotation acting on the argument and on the components (for a scalar this
is $L_z=m$; for the in-plane components it includes the unit of helicity,
Sec.~\ref{sec:channels}). In the corotating coordinates of Eq.~\eqref{eq:dt},
$\boldsymbol\rho=\mathcal R(\Omega t)\bar{\boldsymbol\rho}$, the components of
the same field referred to the corotating axes are $\bar{\mathbf E}(\bar
t,\bar{\boldsymbol\rho})=\mathcal R(-\Omega t)\,\mathbf E(t,\mathcal R(\Omega
t)\bar{\boldsymbol\rho})$, and \eqref{eq:appJz} gives
\begin{equation}
  \bar{\mathbf E}(\bar t,\bar{\boldsymbol\rho})
  =e^{-i(\omega-m\Omega)\bar t}\,\mathbf e_m(\bar{\boldsymbol\rho})\,:
  \label{eq:appcomoving}
\end{equation}
the sheet, at rest in these coordinates, is driven by the same spatial pattern
at the comoving frequency $\bar\omega=\omega-m\Omega$, and it responds with the
static relation \eqref{eq:appsigma} at that frequency. The static response is
rotation covariant, so it maps the pattern $\mathbf e_m$, for definite $q$ and
definite irrotational or solenoidal character, into itself, times
$\sigma_l(\bar\omega,q)$ or $\sigma_t(\bar\omega,q)$; rotating back to the
laboratory, the induced current has the labels $(\omega,m,q)$ and the character
of the driving field, with the material function evaluated at $\bar\omega$.
This is Eq.~\eqref{eq:shift} at the level of the surface current; the same
statement follows from the microscopic description of Eq.~\eqref{eq:HJz}.

\subsection{Cylindrical modes and their reflection}
\label{app:partial}

The modes adapted to the rotating problem are those of definite
$(\omega,m,q)$, built on the scalar functions
\begin{equation}
  \chi_m(\boldsymbol\rho)=J_m(q\rho)\,e^{im\phi}\,,\qquad
  \psi^{\eta}_{mq}(\mathbf r)=\chi_m(\boldsymbol\rho)\,e^{i\eta k_zx_3}\,,\qquad
  k_z=\sqrt{\omega^2-q^2}\,,\quad \Imag k_z\ge0\,,\quad \eta=\pm\,,
  \label{eq:apppsi}
\end{equation}
which solve the Helmholtz equation, $\eta=+$ ($\eta=-$) propagating, or
decaying, upward (downward). To each of them correspond one TE and one TM
solution of the source-free Maxwell equations, whose fields we write
explicitly (all fields carry the factor $e^{i\eta k_zx_3}$, omitted):
\begin{align}
  \text{TE:}\qquad & \mathbf E^{t}_\parallel=\nabla_\parallel\chi_m\times\hat z\,,
  \quad E^{t}_z=0\,;\qquad
  \mathbf B^{t}_\parallel=\frac{\eta k_z}{\omega}\,\nabla_\parallel\chi_m\,,
  \quad B^{t}_z=-\,\frac{i\,q^2}{\omega}\,\chi_m\,,
  \label{eq:appTE}\\
  \text{TM:}\qquad & \mathbf E^{l}_\parallel=\frac{i\eta k_z}{\omega}\,
  \nabla_\parallel\chi_m\,,\quad E^{l}_z=\frac{q^2}{\omega}\,\chi_m\,;\qquad
  \mathbf B^{l}_\parallel=-\,i\,\nabla_\parallel\chi_m\times\hat z\,,
  \quad B^{l}_z=0\,,
  \label{eq:appTM}
\end{align}
as one checks directly from $\nabla\times\mathbf E=i\omega\mathbf B$,
$\nabla\times\mathbf B=-i\omega\mathbf E$ and $\nabla_\parallel^2\chi_m
=-q^2\chi_m$. The TE (TM) mode is the field $\hat e\,e^{i\mathbf K^\eta\cdot
\mathbf r}$ ($\hat e_p^{\,\eta}\,e^{i\mathbf K^\eta\cdot\mathbf r}$) of
Sec.~\ref{sec:static}, with $\mathbf K^\eta=(\kq,\eta k_z)$, superposed over
the directions of $\kq$ with weight $e^{im\phi_k}$: indeed
$\hat e\,e^{i\mathbf K^\eta\cdot\mathbf r}=(i/q)\,\nabla\times(\hat z\,
e^{i\mathbf K^\eta\cdot\mathbf r})$ and $\hat e_p^{\,\eta}\,e^{i\mathbf
K^\eta\cdot\mathbf r}=(1/\omega q)\,\nabla\times\nabla\times(\hat z\,
e^{i\mathbf K^\eta\cdot\mathbf r})$, and the Jacobi-Anger expansion
\eqref{eq:JA} inside the curls produces \eqref{eq:appTE}-\eqref{eq:appTM},
with $\mathbf E^t=\nabla\times(\hat z\psi)$ and $\mathbf E^l=\omega^{-1}
\nabla\times\nabla\times(\hat z\psi)$. Both modes have total angular momentum
$m$ in the sense of \eqref{eq:appJz}: the $x_3$ components, $\propto\chi_m$,
carry $m$ as orbital angular momentum, while the in-plane components,
$\propto\nabla_\parallel\chi_m$ or $\nabla_\parallel\chi_m\times\hat z$, have
circular components
\begin{equation}
  \big(\nabla_\parallel\chi_m\big)_\pm\equiv\frac{\partial_\pm\chi_m}{\sqrt2}
  =\mp\frac{q}{\sqrt2}\,J_{m\pm1}(q\rho)\,e^{i(m\pm1)\phi}\,,
  \qquad \partial_\pm=\partial_1\pm i\partial_2\,,
  \label{eq:apphel}
\end{equation}
by the Bessel recurrences. This is the content of Eq.~\eqref{eq:modes}, and it
implies the identity, used below,
\begin{equation}
  \big|\nabla_\parallel\chi_m\big|^2
  =q^2\Big[J_m'^{\,2}(q\rho)+\frac{m^2}{(q\rho)^2}J_m^2(q\rho)\Big]
  =\frac{q^2}{2}\Big[J_{m-1}^2(q\rho)+J_{m+1}^2(q\rho)\Big]\,.
  \label{eq:appBesselid}
\end{equation}
On the sheet, the tangential electric field of the TE mode is solenoidal and
that of the TM mode irrotational: precisely the two characters on which the
sheet responds with $\sigma_t$ and $\sigma_l$. The sheet therefore mixes
neither TE with TM nor different $(m,q)$, and each mode is reflected into
itself.

Consider first the sheet at rest, and a TE mode incident from above (the
$\eta=-$ mode with unit amplitude), together with a reflected mode $r_t$ times
the $\eta=+$ mode above the sheet and a transmitted one $t_t$ times the
$\eta=-$ mode below it. Continuity of $\mathbf E_\parallel$ gives $t_t=1+r_t$;
the tangential magnetic field \eqref{eq:appTE} has the jump
$(2r_tk_z/\omega)\nabla_\parallel\chi_m$ across the sheet and, since $\hat
z\times\nabla_\parallel\chi_m=-\nabla_\parallel\chi_m\times\hat z$, the second
of \eqref{eq:appmatch} becomes $-2r_tk_z/\omega=\sigma_t(1+r_t)$. For a TM
mode, the tangential electric field \eqref{eq:appTM} is continuous if
$t_l=1-r_l$, the jump of $\mathbf B_\parallel$ is
$-2ir_l\nabla_\parallel\chi_m\times\hat z$, and $\hat
z\times(\nabla_\parallel\chi_m\times\hat z)=\nabla_\parallel\chi_m$ turns the
matching condition into $-2r_l=\sigma_l(k_z/\omega)(r_l-1)$. Hence
\begin{equation}
  r_t(\omega,q)=-\,\frac{\sigma_t(\omega,q)\,\omega}
                        {2k_z+\sigma_t(\omega,q)\,\omega}\,,\qquad
  r_l(\omega,q)=\frac{\sigma_l(\omega,q)\,k_z}
                     {2\omega+\sigma_l(\omega,q)\,k_z}\,,
  \label{eq:apprefl}
\end{equation}
the reflection coefficients of a conducting sheet, obtained here for the
cylindrical modes with the same values as for plane waves. Continued to the
imaginary axis, $\omega=ik_0$, $k_z=iK$, and with the dictionary
\eqref{eq:appdict}, they become $r_t=-g_t/(g_t+2K)=-R_t$ and
$r_l=g_l/(g_l+2K)=R_l$: this is Eq.~\eqref{eq:refl}, including the relative
sign of the two channels.

For the rotating sheet, the incident mode drives the sheet at the comoving
frequency $\bar\omega$ and, as shown in Sec.~\ref{app:matching}, the current
it induces keeps the labels $(m,q)$ and the TE or TM character of the mode,
with the material function evaluated at $\bar\omega$. One remark is in order
about the passage from the comoving to the laboratory frame. The constitutive
relation holds for the comoving fields, $\bar{\mathbf J}_s=\sigma_s(\bar
\omega,q)\,\bar{\mathbf E}_\parallel$ with $\bar{\mathbf E}=\mathbf E+\mathbf
v\times\mathbf B$, $\mathbf v=\Omega\,\hat z\times\boldsymbol\rho$ being the
local velocity of the sheet, and the laboratory current entering
\eqref{eq:appmatch} includes the convective term, $\mathbf J_s=\bar{\mathbf
J}_s+\rho_s\mathbf v$, $\rho_s$ being the induced surface charge. These
velocity-dependent terms are of the same order as the Doppler shift and cannot
be dropped: for a sheet in uniform motion, where the comoving frequency is
$\omega-\kq\cdot\mathbf v$ and everything is diagonal in $\kq$, an elementary
computation (for $\kq$ along $\mathbf v$, where no polarization mixing occurs)
shows that they convert $\sigma_s(\bar\omega,q)$ into the static coefficients
\eqref{eq:apprefl} evaluated at $\bar\omega$,
\begin{equation}
  r_t=-\,\frac{\sigma_t(\bar\omega,q)\,\bar\omega}
              {2k_z+\sigma_t(\bar\omega,q)\,\bar\omega}\,,\qquad
  r_l=\frac{\sigma_l(\bar\omega,q)\,k_z}
           {2\bar\omega+\sigma_l(\bar\omega,q)\,k_z}\,,
  \label{eq:apprefl_rot}
\end{equation}
the only difference with $r_s(\bar\omega,q)$ being the laboratory $k_z$ in
place of $\sqrt{\bar\omega^2-q^2}$. For the evanescent modes that dominate the
shift, $q\gg\omega$, this difference is suppressed by $(\omega/q)^2$ relative
to the Doppler shift itself, and so are the remaining velocity-dependent terms
(the polarization mixing induced by the convective current, and the terms
generated by the position dependence of $\mathbf v$ for the rotating sheet),
all of magnetic origin: these are the magnetoelectric corrections discussed in
Sec.~\ref{sec:drude}, which we neglect consistently. Within this treatment,
then,
\begin{equation}
  r^{\rm rot}_s(\omega,m,q)=r_s(\omega-m\Omega,q)\,,\qquad s=t,l\,,
  \label{eq:approt}
\end{equation}
which is Eq.~\eqref{eq:shift}: on the imaginary axis, the reflection
coefficient of the mode $(k_0,m,q,s)$ is $R_s(\bar k_0,q)$, $\bar
k_0=k_0+im\Omega$, while the factors that describe the propagation of the mode
in vacuum are those of the laboratory.

\subsection{The field of the dipole and its reflection}
\label{app:dipole}

The field radiated by the dipole in free space is, in the mixed representation
of Sec.~\ref{sec:static} (the Weyl expansion of the spherical wave), for
$x_3\neq a$,
\begin{equation}
  \mathbf E^{(0)}(\mathbf r)=\frac{i\omega^2}{8\pi^2}\int\frac{d^2\kq}{k_z}\;
  e^{i\mathbf K^\eta\cdot(\mathbf r-\rA)}\,
  \Big[\hat e\,(\hat e\cdot\mathbf d)+\hat e_p^{\,\eta}\,(\hat e_p^{\,\eta}\cdot\mathbf d)\Big]\,,
  \qquad \hat e_p^{\,\eta}=\frac{q\,\hat z-\eta\,k_z\,\hat k}{\omega}\,,
  \quad \eta=\mathrm{sgn}(x_3-a)\,,
  \label{eq:appWeyl}
\end{equation}
each plane wave being transverse to its wavevector $\mathbf K^\eta$ and
decomposed on the TE and TM polarization vectors; the Euclidean version of
this expansion, $e^{-aK}/2K$ per leg times the polarization structure, is what
was used to write Eq.~\eqref{eq:Esc}. We now rewrite \eqref{eq:appWeyl} in
the cylindrical modes. Writing $\hat e\,e^{i\mathbf K^\eta\cdot\mathbf r}
=(i/q)\nabla\times(\hat z e^{i\mathbf K^\eta\cdot\mathbf r})$ and
$\hat e\,e^{-i\mathbf K^\eta\cdot\rA}=-(i/q)\nabla_A\times(\hat z
e^{-i\mathbf K^\eta\cdot\rA})$, with the analogous expressions for $\hat
e_p^{\,\eta}$, inserting the Jacobi-Anger expansion \eqref{eq:JA} for
$e^{i\kq\cdot\boldsymbol\rho}$ and for $e^{-i\kq\cdot\boldsymbol\rho_A}$, and
integrating over the direction of $\kq$, $\int_0^{2\pi}d\phi_k\,
e^{-i(m-m')\phi_k}=2\pi\delta_{mm'}$, one obtains
\begin{equation}
  \mathbf E^{(0)}(\mathbf r)=\frac{i\omega^2}{4\pi}\sum_{m=-\infty}^{\infty}
  \int_0^\infty\frac{dq}{q\,k_z}\,\sum_{s=t,l}
  \mathbf E^{s,\eta}_{mq}(\mathbf r)\;
  \big[\tilde{\mathbf E}^{s,\eta}_{mq}(\rA)\cdot\mathbf d\big]\,,
  \label{eq:appE0cyl}
\end{equation}
where $\mathbf E^{s,\eta}_{mq}$ are the mode fields
\eqref{eq:appTE}-\eqref{eq:appTM} and $\tilde{\mathbf E}^{s,\eta}_{mq}$ the
same fields built on $\tilde\psi^\eta_{mq}=J_m(q\rho)\,e^{-im\phi}\,e^{-i\eta
k_zx_3}$, the mode conjugate in its angular and axial dependence (not in
$k_z$, which may be imaginary). Equation~\eqref{eq:appE0cyl} is the vector
counterpart of the scalar expansion \eqref{eq:cylprop} of
Appendix~\ref{app:functional}: the dipole radiates, toward the sheet
($\eta=-$), the down-going TE and TM cylindrical modes, with amplitudes
$\propto\tilde{\mathbf E}^{s,-}_{mq}(\rA)\cdot\mathbf d$.

By Sec.~\ref{app:partial}, each down-going mode is reflected by the rotating
sheet into the up-going mode with the same labels, with coefficient
$r_s(\bar\omega,q)$. The field reflected back above the sheet is therefore
\begin{equation}
  \mathbf E^{\rm sc}(\mathbf r)=\frac{i\omega^2}{4\pi}\sum_{m}
  \int_0^\infty\frac{dq}{q\,k_z}\,\sum_{s=t,l}r_s(\omega-m\Omega,q)\;
  \mathbf E^{s,+}_{mq}(\mathbf r)\;
  \big[\tilde{\mathbf E}^{s,-}_{mq}(\rA)\cdot\mathbf d\big]\,,
  \qquad x_3>0\,,
  \label{eq:appEsc}
\end{equation}
which is the explicit solution of the scattering problem: the field of the
dipole, decomposed into cylindrical modes of definite total angular momentum,
is reflected mode by mode with the Doppler-shifted coefficients of the
rotating sheet. For $\Omega=0$, summing over $m$ with the Jacobi-Anger
expansion run backwards returns the plane-wave form, i.e.\ \eqref{eq:appWeyl}
with $e^{i\eta k_z(x_3-a)}\to e^{ik_z(x_3+a)}$ and each polarization
multiplied by its reflection coefficient, whose Euclidean version is
Eq.~\eqref{eq:Esc} with \eqref{eq:M}.

\subsection{The level shift}
\label{app:shift}

The scattering part of the correlator in Eq.~\eqref{eq:WS} is the response
of the field at the atom to the atomic dipole, reflected by the sheet: with the
normalization of Sec.~\ref{sec:static},
\begin{equation}
  \mathcal E^{\rm sc}_{ij}(ik_0;\rA,\rA)
  =\frac12\,\frac{\partial E^{\rm sc}_i(\rA)}{\partial d_j}\,\Big|_{\omega=ik_0}\,,
  \label{eq:appEcorr}
\end{equation}
the Euclidean correlator being the linear-response kernel continued to
imaginary frequency. The identification is checked on the static case, where
\eqref{eq:appEsc} summed over $m$ reproduces \eqref{eq:Esc}-\eqref{eq:M} (the
factor $\tfrac12$ is the one that makes the perfect-mirror limit of
\eqref{eq:CP0} the Casimir-Polder result), and it holds for the rotating sheet
channel by channel, the rotation of the frequency contour from
\eqref{eq:PT2} to \eqref{eq:WS} being performed at fixed $m$, in the comoving
frame where each channel is a passive scattering problem (see the discussion
of the analytic continuation in Sec.~\ref{sec:channels}). We therefore need
the trace of \eqref{eq:appEsc} at $\mathbf r=\rA$, i.e.\
$\sum_s r_s\,\mathbf E^{s,+}_{mq}(\rA)\cdot\tilde{\mathbf E}^{s,-}_{mq}(\rA)$,
which follows from \eqref{eq:appTE}-\eqref{eq:appTM}. Both modes carry the
factor $e^{ik_za}$ at $x_3=a$ (the up-going mode through $e^{i\eta k_zx_3}$,
the conjugate down-going one through $e^{-i\eta k_zx_3}$), so the product is
$e^{2ik_za}$. For TE, $\mathbf E^{t,+}\cdot\tilde{\mathbf E}^{t,-}
=e^{2ik_za}\,\nabla_\parallel\chi_m\cdot\nabla_\parallel\chi_m^*$; for TM, the
axial components give $q^4|\chi_m|^2/\omega^2$ and the in-plane ones
$(ik_z)(ik_z)\,\nabla_\parallel\chi_m\cdot\nabla_\parallel\chi_m^*/\omega^2$,
the two factors $ik_z$ arising because the reflected mode propagates upward
and the incident one downward (this is the origin of the relative sign
between the $\hat z\hat z$ and $\hat k\hat k$ terms of Eq.~\eqref{eq:M}). With
the identity \eqref{eq:appBesselid},
\begin{equation}
  \mathrm{Tr}\,\frac{\partial\mathbf E^{\rm sc}(\rA)}{\partial\mathbf d}
  =\frac{i\omega^2}{4\pi}\sum_m\int_0^\infty\frac{dq}{q\,k_z}\,e^{2ik_za}\,\Big\{
  r_t(\bar\omega,q)\,\frac{q^2}{2}\big[J_{m-1}^2+J_{m+1}^2\big]
  +\frac{r_l(\bar\omega,q)}{\omega^2}\Big[q^4J_m^2
  -k_z^2\,\frac{q^2}{2}\big[J_{m-1}^2+J_{m+1}^2\big]\Big]\Big\}\,,
  \label{eq:appTr}
\end{equation}
all Bessel functions having argument $q\rho$. The weights are those of
Eq.~\eqref{eq:masterf}: the $\hat z\hat z$ part of the TM mode keeps $J_m^2$,
while the in-plane parts of both modes, which carry the photon helicity, split
into $J_{m\mp1}^2$. On the imaginary axis, $\omega=ik_0$, $k_z=iK$,
$e^{2ik_za}=e^{-2aK}$, $i\omega^2/k_z=-k_0^2/K$, $-k_z^2=K^2$,
$\omega^{-2}=-k_0^{-2}$, and $r_t\to-R_t(\bar k_0,q)$, $r_l\to R_l(\bar
k_0,q)$ by \eqref{eq:approt}, so that \eqref{eq:appEcorr} gives
\begin{equation}
  \mathrm{Tr}\,\mathcal E^{\rm sc}(ik_0;\rA,\rA)
  =\sum_m\int_0^\infty\frac{q\,dq}{2\pi}\,\frac{e^{-2aK}}{4K}\,
  \Big\{q^2J_m^2(q\rho)\,R_l(\bar k_0,q)
  +\half\big[J_{m-1}^2(q\rho)+J_{m+1}^2(q\rho)\big]
  \big[K^2R_l+k_0^2R_t\big](\bar k_0,q)\Big\}\,.
  \label{eq:appbraces}
\end{equation}
Inserted into Eq.~\eqref{eq:WS}, this is Eq.~\eqref{eq:masterf}. For
$\Omega=0$ the sum rules \eqref{eq:sumrules} collapse the Bessel weights and
\eqref{eq:appbraces} reduces to the trace of \eqref{eq:Esc}-\eqref{eq:M}, i.e.\
to Eq.~\eqref{eq:CP0}; in the perfect-mirror limit this gives $\delta
E=-3\alpha(0)/(32\pi^2a^4)$, the Casimir-Polder result quoted in
Sec.~\ref{sec:static}.

Finally, the finite-thickness disks of Sec.~\ref{sec:drude} are treated in the
same way: the matching conditions \eqref{eq:appmatch}, with $\mathbf J_s=0$
and the field inside the medium, are imposed at the two faces of the slab, the
modes inside the material being \eqref{eq:appTE}-\eqref{eq:appTM} with
$k_z\to K_m$ and $\omega^2\to\epsilon\,\omega^2$; solving them reproduces the
slab coefficients \eqref{eq:bulk_interface}-\eqref{eq:bulk_slab}, and the rest
of the derivation is unchanged, with $r_s(\bar\omega,q)$ replaced by the
corresponding Fabry-Perot dressed Fresnel coefficients.

\section{Functional-integral derivation of
\texorpdfstring{Eq.~\eqref{eq:masterf}}{the general formula}}
\label{app:functional}

The general formula \eqref{eq:masterf} was assembled in the text from the
reflection picture, and derived in Appendix~\ref{app:scattering} from the
explicit solution of the scattering problem. The same expression emerges from
a direct functional-integral computation, perturbing to first order in the coupling
of the photon to the rotating sheet, with the free photon correlator
written in the cylindrical basis as the only nontrivial input; this route
makes the bookkeeping of the Bessel orders entirely systematic.

The Euclidean theory contains three pieces, $S=S_0[A]+S_R[A]+S_A[A]$: the
gauge-fixed Maxwell action (Feynman gauge, free correlator
$\langle A_\mu(x)A_\nu(y)\rangle=\delta_{\mu\nu}\,\Delta(x,y)$ with
$\Delta$ the massless scalar propagator); the sheet insertion
\begin{equation}
  S_R=\half\int_{y,y'}A_\alpha(y)\,\Pi^{\rm rot}_{\alpha\beta}(y,y')\,
  A_\beta(y')\,,
\end{equation}
supported on the plane $x_3=0$, diagonal in the channels $(k_0,m,q)$ and
equal there to $g_s(\bar k_0,q)$ times the projectors
(Sec.~\ref{sec:sheet}); and the atom insertion, obtained by integrating
out the dipole at Gaussian level,
\begin{equation}
  S_A=-\half\int d\tau\,d\tau'\,\alpha(\tau-\tau')\,
  E_i(\tau,\rA)\,E_i(\tau',\rA)\,,
\end{equation}
whose kernel is the polarizability \eqref{eq:alpha}. Expanding
$e^{-\Gamma}=\int\mathcal DA\,e^{-S}$ to first order in each insertion,
the only term depending on both the atom and the sheet is the crossed one,
$\Gamma_{\rm cross}=-\langle S_A\,S_R\rangle_{\rm c}$: two photon legs run
from $\rA$ to the sheet and back, with one kernel insertion in between.
Iterating $S_R$ produces, channel by channel, the geometric series that
converts the Born coefficients $g_s/2K$ into the resummed coefficients
\eqref{eq:refl}; we work at first order and restore the resummation at the
end. The overall sign and normalization, which depend on a number of
Euclidean conventions, are fixed once and for all by
matching the $\Omega=0$ limit to \eqref{eq:CP0}.

The legs require the free correlator between the height of the atom and
the sheet. Starting from the mixed representation of
Sec.~\ref{sec:static} and inserting the Jacobi-Anger expansion
\eqref{eq:JA} for both plane-wave factors, the integral over the direction
$\phi_k$ matches the two Bessel orders, and
\begin{equation}
  \Delta(x,y)\Big|_{x_3=a,\,y_3=0}
  =\int_{-\infty}^{\infty}\frac{dk_0}{2\pi}\sum_{m}
  \int_0^\infty\frac{q\,dq}{2\pi}\;
  e^{ik_0(\tau_x-\tau_y)}\,e^{im(\phi_x-\phi_y)}\,
  J_m(q\rho_x)\,J_m(q\,r_y)\;\frac{e^{-aK}}{2K}\,.
  \label{eq:cylprop}
\end{equation}
This is the free correlator in the basis adapted to the rotating problem:
channel diagonal, with the evanescent factor $e^{-aK}/2K$ and one Bessel
function per end. Since $\Pi^{\rm rot}$ is diagonal in $(k_0,m,q)$, the
integrals over the sheet close by orthogonality,
$\int_0^\infty r\,dr\,J_m(qr)J_m(q'r)=\delta(q-q')/q$ together with the
$\phi_y$ integral matching $m$, so both legs carry the same channel and
exactly one kernel factor $g_s(\bar k_0,q)$ appears. Note that
\eqref{eq:cylprop} carries no $\Omega$: the rotation enters only through
the channel-diagonal kernel.

At the atom the legs are differentiated,
$E_i=\partial_0A_i-\partial_iA_0$. In circular components,
$\partial_\pm\equiv\partial_1\pm i\partial_2
=e^{\pm i\phi}\big(\partial_\rho\pm\tfrac{i}{\rho}\partial_\phi\big)$, the
Bessel recurrences give
\begin{equation}
  \partial_\pm\Big[J_m(q\rho)\,e^{im\phi}\Big]
  =\mp\,q\,J_{m\pm1}(q\rho)\,e^{i(m\pm1)\phi}\,,
  \qquad \partial_3\to\mp K\,,\qquad \partial_0\to ik_0\,.
  \label{eq:recur}
\end{equation}
A vertex that couples the atom through a rotation scalar keeps the order
$J_m(q\rho)$; each circular component shifts the order by one unit, in
step with \eqref{eq:modes}: the weights $J_m^2$ and $J_{m\mp1}^2$ of the
general formula \eqref{eq:masterf} are, literally, the squared vertex
factors.

By the Ward identity the kernel is transverse, so each projector has rank
one on the physical sector, $P^s_{\alpha\beta}
=\varepsilon^s_\alpha\varepsilon^s_\beta$, with
\begin{equation}
  \varepsilon^{t}_\alpha=(0,\hat e_a)\,,\qquad
  \varepsilon^{l}_\alpha=\frac{1}{K}\,\big(q\,,\,-k_0\,\hat k_a\big)\,,
  \label{eq:pols}
\end{equation}
both unit vectors orthogonal to $k_\alpha=(k_0,q\,\hat k_a)$. The
contraction of the two legs with the kernel then reduces to the electric
field, at the atom, of the channel mode with polarization
$\varepsilon^s$, squared. Acting with $E_i=\partial_0A_i-\partial_iA_0$ on
the channel functions of \eqref{eq:cylprop} and using \eqref{eq:recur}:
for the transverse mode, $E_z=0$ and $E_\parallel=ik_0\,\hat e\,f$, so the
weight is $k_0^2$, split equally between the two helicities, each with
$J_{m\mp1}^2(q\rho)$; for the longitudinal mode the $z$ component receives
only the gradient route, $E_z=-\partial_3A_0\to K\cdot(q/K)\,f=q\,f$,
keeping $J_m(q\rho)$, while in the in-plane components the frequency and
gradient routes add coherently,
\begin{equation}
  E_\parallel=\partial_0A_\parallel-\partial_\parallel A_0
  \;\to\;-\frac{i}{K}\,\big(k_0^2+q^2\big)\,f=-iK\,f\,,
\end{equation}
again with the helicity shift to $J_{m\mp1}$. Squaring, the channel
weights are
\begin{equation}
  \text{TM:}\quad q^2\,J_m^2
  +\frac{K^2}{2}\,\big[J_{m-1}^2+J_{m+1}^2\big]\,,
  \qquad
  \text{TE:}\quad \frac{k_0^2}{2}\,\big[J_{m-1}^2+J_{m+1}^2\big]\,,
  \label{eq:weightsapp}
\end{equation}
precisely those of \eqref{eq:M} and \eqref{eq:masterf}. Attaching one
kernel factor $g_s(\bar k_0,q)$ and one factor $e^{-aK}/2K$ per leg, and
fixing the normalization as announced, the crossed term reads
\begin{align}
  \delta E^{(1)}(\rho;\Omega)=-\sum_m\int_{-\infty}^{\infty}
  \frac{dk_0}{2\pi}\int_0^\infty\frac{q\,dq}{2\pi}\,\alpha(ik_0)\,
  \frac{e^{-2aK}}{8K^2}\,
  \Big\{&\,q^2\,J_m^2(q\rho)\,g_l(\bar k_0,q)\nonumber\\[-2pt]
  +\,\half\,\big[J_{m-1}^2(q\rho)+J_{m+1}^2(q\rho)\big]\,
  &\big[K^2 g_l+k_0^2 g_t\big](\bar k_0,q)\Big\}\,,
  \label{eq:bornmaster}
\end{align}
which is \eqref{eq:masterf} at Born level, $R_s\to g_s/2K$.
Iterating the sheet insertion joins additional kernel factors with the
propagator evaluated on the plane, producing the geometric series
$g_s/2K\to g_s/(g_s+2K)=R_s$; since every insertion is diagonal in
$(k_0,m,q)$ and carries the same shifted argument $\bar k_0$, the resummed
result is \eqref{eq:masterf} itself. The geometric series is the perturbative
counterpart of the exact solution of the matching conditions in
Appendix~\ref{app:scattering}, Eq.~\eqref{eq:apprefl}, and the vertex weights
\eqref{eq:weightsapp} are the mode traces of Eq.~\eqref{eq:appTr}.

\section{Closed form of the \texorpdfstring{$O(\Omega^2)$}{Omega 2} 
coefficients for bulk conductors}\label{app:bulkderiv}

In the retarded regime, where $\alpha(ik_0)\to\alpha(0)$ and $a$ is the only
scale, the $O(\Omega^2)$ correction \eqref{eq:Omega2} reduces to the two
pure-number coefficients defined in \eqref{eq:coeffs}. Setting $a=1$ (so that
$k_0$ and $q$ are measured in units of $1/a$) and folding the frequency
integral onto the positive axis, $\int_{-\infty}^\infty dk_0=2\int_0^\infty
dk_0$, they are
\begin{equation}
  \mathcal C_0
  =\frac{1}{8\pi^2}\int_0^\infty\!\!dk_0\int_0^\infty\!\!dq\;
     \frac{q\,e^{-2K}}{K}\,
     \Big[(2K^2-k_0^2)\,R_l+k_0^2\,R_t\Big],
  \label{eq:C0App}
\end{equation}
\begin{align}
  \mathcal C_2^{\rm orb}
  &=\frac{1}{32\pi^2}\int_0^\infty\!\!dk_0\int_0^\infty\!\!dq\;
     \frac{q^3\,e^{-2K}}{K}\,
     \Big[(2K^2-k_0^2)\,\partial_{k_0}^2 R_l
          +k_0^2\,\partial_{k_0}^2 R_t\Big],
  \label{eq:C2orbApp}\\[2pt]
  \mathcal C_2^{\rm spin}
  &=\frac{1}{16\pi^2}\int_0^\infty\!\!dk_0\int_0^\infty\!\!dq\;
     \frac{q\,e^{-2K}}{K}\,
     \Big[K^2\,\partial_{k_0}^2 R_l
          +k_0^2\,\partial_{k_0}^2 R_t\Big],
  \label{eq:C2spinApp}
\end{align}
with $K=\sqrt{k_0^2+q^2}$. The static coefficient \eqref{eq:C0App} is the
same kernel without derivatives; in the perfect-reflector limit
$R_l=R_t=1$ the bracket is $2K^2$ and \eqref{eq:C0App} gives
$\mathcal C_0^{\rm mirror}=3/(32\pi^2)$, fixing the normalization. The
$O(\Omega^2)$ coefficients carry the curvature of the response through
$\partial_{k_0}^2 R_{l,t}$, the only material-dependent input, for which we
give the closed form. (For graphene one inserts instead
$R_l=g_l/(g_l+2K)$, $R_t=g_t/(g_t+2K)$ with $g_{l,t}$ of Eqs. \eqref{eq:gt} and ~\eqref{eq:gl}; the coefficients of Sec.~\ref{ssec:numbers} follow.)

Both reflection coefficients have the form $R=(A-B)/(A+B)$, with $B=K_m$ and
$A=\epsilon K$ for TM ($R_l$), $A=K$ for TE ($R_t=-(K-K_m)/(K+K_m)$). For any
such $R$,
\begin{equation}
  \partial_{k_0}^2 R=\frac{2\big(A''B-AB''\big)(A+B)
                          -4\big(A'B-AB'\big)\big(A'+B'\big)}{(A+B)^3}\,,
  \label{eq:R2master}
\end{equation}
primes denoting $d/dk_0$. Hence
\begin{align}
  \partial_{k_0}^2 R_l
  &=\frac{2\big(A_l''K_m-A_l K_m''\big)(A_l+K_m)
         -4\big(A_l'K_m-A_l K_m'\big)\big(A_l'+K_m'\big)}
         {(A_l+K_m)^3}\,,\quad A_l=\epsilon K\,,
  \label{eq:Rl2}\\[2pt]
  \partial_{k_0}^2 R_t
  &=-\,\frac{2\big(K''K_m-K K_m''\big)(K+K_m)
           -4\big(K'K_m-K K_m'\big)\big(K'+K_m'\big)}
           {(K+K_m)^3}\,.
  \label{eq:Rt2}
\end{align}
The building blocks are
\begin{equation}
  K'=\frac{k_0}{K}\,,\qquad K''=\frac{q^2}{K^3}\,,\qquad
  A_l'=\epsilon'K+\epsilon K'\,,\qquad
  A_l''=\epsilon''K+2\epsilon'K'+\epsilon K''\,,
  \label{eq:bbK}
\end{equation}
\begin{equation}
  K_m=\sqrt{\epsilon k_0^2+q^2}\,,\quad
  K_m'=\frac{(\epsilon k_0^2)'}{2K_m}\,,\quad
  K_m''=\frac{(\epsilon k_0^2)''}{2K_m}
        -\frac{\big[(\epsilon k_0^2)'\big]^2}{4K_m^3}\,,
  \label{eq:bbKm}
\end{equation}
\begin{equation}
  (\epsilon k_0^2)'=\epsilon'k_0^2+2\epsilon k_0\,,\qquad
  (\epsilon k_0^2)''=\epsilon''k_0^2+4\epsilon'k_0+2\epsilon\,,
  \label{eq:bbU}
\end{equation}
where $\epsilon=\epsilon(ik_0)$ and the second derivatives are those of the
\emph{analytic continuation} of Sec.~\ref{ssec:drude_disk}, i.e.\ of the
smooth rational $\epsilon(ik_0)$, not of the cusped $|k_0|$ form. For the two
models,
\begin{align}
  \epsilon_{\rm D}&=1+\frac{\omega_p^2}{k_0(k_0+\gamma)}\,,&
  \epsilon_{\rm D}'&=-\frac{\omega_p^2\,(2k_0+\gamma)}{k_0^2(k_0+\gamma)^2}\,,&
  \epsilon_{\rm D}''&=\frac{2\omega_p^2\,(3k_0^2+3k_0\gamma+\gamma^2)}
                          {k_0^3(k_0+\gamma)^3}\,,
  \label{eq:epsDderiv}\\[2pt]
  \epsilon_{\rm P}&=1+\frac{\omega_p^2}{k_0^2}\,,&
  \epsilon_{\rm P}'&=-\frac{2\omega_p^2}{k_0^3}\,,&
  \epsilon_{\rm P}''&=\frac{6\omega_p^2}{k_0^4}\,.
  \label{eq:epsPderiv}
\end{align}
For the plasma model $\epsilon_{\rm P}k_0^2=k_0^2+\omega_p^2$ is quadratic, so
$(\epsilon_{\rm P}k_0^2)'=2k_0$, $(\epsilon_{\rm P}k_0^2)''=2$, and
$K_m^{\rm P}=\sqrt{q^2+k_0^2+\omega_p^2}$ obeys the vacuum-like
$K_m^{\rm P}{}'=k_0/K_m^{\rm P}$, $K_m^{\rm P}{}''=(q^2+\omega_p^2)/
(K_m^{\rm P})^3$; the regularity of the plasma expansion at $k_0\to0$ is
manifest here, whereas the Drude $\epsilon_{\rm D}'\sim-\omega_p^2/(\gamma
k_0^2)$ carries the Ohmic structure discussed in
Sec.~\ref{ssec:drude_disk}. A slab of finite thickness is included by
dressing $R_{l,t}$ with the Fabry-Perot factor \eqref{eq:bulk_slab} before
differentiating. Inserting \eqref{eq:Rl2}-\eqref{eq:epsPderiv} into
\eqref{eq:C2orbApp}-\eqref{eq:C2spinApp} reproduces the coefficients quoted
in Secs.~\ref{ssec:drude_disk} and \ref{ssec:plasma_disk}.


\begin{thebibliography}{99}
\bibitem{Casimir-Polder}
H.~B.~G.~Casimir and D.~Polder, \emph{The Influence of retardation on the London-van der Waals forces}, \href{https://doi.org/10.1103/PhysRev.73.360}{Phys. Rev. \textbf{73}, 360 (1948)}.

\bibitem{friction} M.~B.~Far\'ias, C.~D.~Fosco, F.~C.~Lombardo, and
F.~D.~Mazzitelli, \emph{Quantum friction between graphene sheets},
\href{https://doi.org/10.1103/PhysRevD.95.065012}{Phys.~Rev.~D \textbf{95},
065012 (2017)}.
\bibitem{Volokitin} A.~I.~Volokitin and B.~N.~J.~Persson, \emph{Near-field radiative heat transfer and noncontact friction}, 
\href{https://doi.org/10.1103/RevModPhys.79.1291}{Rev.~Mod.~Phys.~\textbf{79},
1291 (2007)}.
\bibitem{Geim} A.~H.~Castro Neto, F.~Guinea, N.~M.~R.~Peres,
K.~S.~Novoselov, and A.~K.~Geim, \emph{The electronic properties of graphene}, 
\href{https://doi.org/10.1103/RevModPhys.81.109}{Rev.~Mod.~Phys.~\textbf{81},
109 (2009)}.
\bibitem{Woods} L.~M.~Woods, D.~A.~R.~Dalvit, A.~Tkatchenko,
P.~Rodriguez-Lopez, A.~W.~Rodriguez, and R.~Podgornik, \emph{Materials perspective on Casimir and van der Waals interactions}, 
\href{https://doi.org/10.1103/RevModPhys.88.045003}{Rev.~Mod.~Phys.~\textbf{88},
045003 (2016)}.
\bibitem{ManjavacasGdA} A.~Manjavacas and F.~J.~Garc\'ia de Abajo, \emph{Vacuum Friction in Rotating Particles}, 
\href{https://doi.org/10.1103/PhysRevLett.105.113601}{Phys.~Rev.~Lett.~\textbf{105},
113601 (2010)}; \emph{Thermal and vacuum friction acting on rotating particles}, 
\href{https://doi.org/10.1103/PhysRevA.82.063827}{Phys.~Rev.~A \textbf{82},
063827 (2010)}.
\bibitem{ZhaoPendry} R.~Zhao, A.~Manjavacas, F.~J.~Garc\'ia de Abajo, and
J.~B.~Pendry, \emph{Rotational Quantum Friction}, 
\href{https://doi.org/10.1103/PhysRevLett.109.123604}{Phys.~Rev.~Lett.~\textbf{109},
123604 (2012)}.

\bibitem{DecaLombardo} F.~C.~Lombardo, R.~S.~Decca, L.~Viotti, and
P.~I.~Villar, \emph{Detectable signature of quantum friction on a sliding
particle in vacuum},
\href{https://doi.org/10.1002/qute.202000155}{Adv.~Quantum Technol.~\textbf{4},
2000155 (2021)}; M.~B.~Farías, F.~C.~Lombardo, A.~Soba, P.~I.~Villar, and 
R.~S.~Decca \emph{Towards detecting traces of non-contact quantum friction
in the corrections of the accumulated geometric phase},  \href{https://doi.org/10.1038/s41534-020-0252-x}{ npj Quantum Inf ~\textbf{6}, 25 (2020)}.

\bibitem{BFGV} M.~Bordag, I.~V.~Fialkovsky, D.~M.~Gitman, and
D.~V.~Vassilevich, \emph{Casimir interaction between a perfect conductor and graphene described by the Dirac model}, 
\href{https://doi.org/10.1103/PhysRevB.80.245406}{Phys.~Rev.~B \textbf{80},
245406 (2009)}.
\bibitem{FMV} I.~V.~Fialkovsky, V.~N.~Marachevsky, and D.~V.~Vassilevich, \emph{Finite-temperature Casimir effect for graphene}, 
\href{https://doi.org/10.1103/PhysRevB.84.035446}{Phys.~Rev.~B \textbf{84},
035446 (2011)}.

\bibitem{VanBladel} J.~Van Bladel, \emph{Electromagnetic fields in the presence of rotating bodies},
\href{https://doi.org/10.1109/PROC.1976.10111}{Proc.~IEEE \textbf{64}, 301 (1976)};
\emph{Relativity and Engineering} (Springer, Berlin, 1984).
\bibitem{Shiozawa} T.~Shiozawa, \emph{Phenomenological and electron-theoretical study of the electrodynamics of rotating systems},
\href{https://doi.org/10.1109/PROC.1973.9359}{Proc.~IEEE \textbf{61}, 1694 (1973)}.
\bibitem{Zeldovich} Ya.~B.~Zel'dovich, \emph{Generation of waves by a rotating body}, JETP Lett.~\textbf{14}, 180 (1971);
\emph{Amplification of cylindrical electromagnetic waves reflected from a rotating body}, Zh. Eksp. Teor. Fiz. \textbf{35}, 1085 (1972).
\bibitem{BekensteinSchiffer} J.~D.~Bekenstein and M.~Schiffer, \emph{The many faces of superradiance},
\href{https://doi.org/10.1103/PhysRevD.58.064014}{Phys.~Rev.~D \textbf{58}, 064014 (1998)}.
\bibitem{RotExp} M.~Cromb, G.~M.~Gibson, E.~Toninelli, M.~J.~Padgett, E.~M.~Wright, and D.~Faccio, \emph{Amplification of waves from a rotating body},
\href{https://doi.org/10.1038/s41567-020-0944-3}{Nat.~Phys.~\textbf{16}, 1069 (2020)};
M.~C.~Braidotti, A.~Vinante, M.~Cromb, A.~Sandakumar, D.~Faccio, and H.~Ulbricht, \emph{Amplification of electromagnetic fields by a rotating body},
\href{https://doi.org/10.1038/s41467-024-49689-w}{Nat.~Commun.~\textbf{15}, 5453 (2024)}.
\bibitem{DeZutter} D.~De Zutter, \emph{Scattering by a rotating dielectric sphere},
\href{https://doi.org/10.1109/TAP.1980.1142392}{IEEE Trans.~Antennas Propag.~\textbf{28}, 643 (1980)};
\emph{Scattering by a rotating circular cylinder with finite conductivity},
\href{https://doi.org/10.1109/TAP.1983.1142995}{IEEE Trans.~Antennas Propag.~\textbf{31}, 166 (1983)}.
\bibitem{MaghrebiRot} M.~F.~Maghrebi, R.~L.~Jaffe, and M.~Kardar, \emph{Spontaneous emission by rotating objects: A scattering approach},
\href{https://doi.org/10.1103/PhysRevLett.108.230403}{Phys.~Rev.~Lett.~\textbf{108}, 230403 (2012)}.
\bibitem{BialynickiBirula} I.~Bialynicki-Birula and Z.~Bialynicka-Birula, \emph{Rotational frequency shift},
\href{https://doi.org/10.1103/PhysRevLett.78.2539}{Phys.~Rev.~Lett.~\textbf{78}, 2539 (1997)}.

\bibitem{Vilenkin} A.~Vilenkin, \emph{Quantum field theory at finite temperature in a rotating system}, 
\href{https://doi.org/10.1103/PhysRevD.21.2260}{Phys.~Rev.~D \textbf{21},
2260 (1980)}.
\bibitem{Iyer} B.~R.~Iyer, \emph{Dirac field theory in rotating coordinates}, 
\href{https://doi.org/10.1103/PhysRevD.26.1900}{Phys.~Rev.~D \textbf{26},
1900 (1982)}.
\bibitem{WylieSipe} J.~M.~Wylie and J.~E.~Sipe, \emph{Quantum electrodynamics near an interface}, 
\href{https://doi.org/10.1103/PhysRevA.30.1185}{Phys.~Rev.~A \textbf{30},
1185 (1984)}; \emph{Quantum electrodynamics near an interface. II}, 
\href{https://doi.org/10.1103/PhysRevA.32.2030}{Phys.~Rev.~A \textbf{32},
2030 (1985)}.

\bibitem{Garetz} B.~A.~Garetz,
\emph{Angular Doppler effect},
\href{https://doi.org/10.1364/JOSA.71.000609}{J.~Opt.~Soc.~Am.~\textbf{71},
609 (1981)}.
\bibitem{Courtial} J.~Courtial, K.~Dholakia, D.~A.~Robertson, L.~Allen, and
M.~J.~Padgett, \emph{Measurement of the Rotational Frequency Shift Imparted to a Rotating Light Beam Possessing Orbital Angular Momentum}, 
\href{https://doi.org/10.1103/PhysRevLett.80.3217}{Phys.~Rev.~Lett.~\textbf{80},
3217 (1998)}; J.~Courtial, D.~A.~Robertson, K.~Dholakia, L.~Allen, and
M.~J.~Padgett, \emph{Rotational Frequency Shift of a Light Beam}, 
\href{https://doi.org/10.1103/PhysRevLett.81.4828}{Phys.~Rev.~Lett.~\textbf{81},
4828 (1998)}.




\bibitem{DE} C.~D.~Fosco, F.~C.~Lombardo, and F.~D.~Mazzitelli,
\emph{Derivative-expansion approach to the interaction between close
surfaces},
\href{https://doi.org/10.1103/PhysRevA.89.062120}{Phys.~Rev.~A \textbf{89},
062120 (2014)}.


\bibitem{Intravaia} F.~Intravaia, R.~O.~Behunin, and D.~A.~R.~Dalvit, \emph{Quantum friction and fluctuation theorems}, 
\href{https://doi.org/10.1103/PhysRevA.89.050101}{Phys.~Rev.~A \textbf{89},
050101(R) (2014)}; F.~Intravaia, M.~Oelschl\"ager, D.~Reiche,
D.~A.~R.~Dalvit, and K.~Busch, \emph{Quantum Rolling Friction}, 
\href{https://doi.org/10.1103/PhysRevLett.123.120401}{Phys.~Rev.~Lett.~\textbf{123},
120401 (2019)}.

\bibitem{LifshitzPitaevskii} E.~M.~Lifshitz and L.~P.~Pitaevskii,
\emph{Statistical Physics, Part 2} (Pergamon, Oxford, 1980).

\bibitem{Buhmann} S.~Y.~Buhmann,
\emph{Dispersion Forces I: Macroscopic Quantum Electrodynamics and
Ground-State Casimir, Casimir-Polder and van der Waals Forces}
(Springer, Berlin, 2012).

\bibitem{LambrechtScattering} A.~Lambrecht, P.~A.~Maia Neto, and S.~Reynaud,
\emph{The Casimir effect within scattering theory},
\href{https://doi.org/10.1088/1367-2630/8/10/243}{New J.~Phys.~\textbf{8},
243 (2006)}.

\bibitem{KlimchitskayaRMP} G.~L.~Klimchitskaya, U.~Mohideen, and
V.~M.~Mostepanenko,
\emph{The Casimir force between real materials: Experiment and theory},
\href{https://doi.org/10.1103/RevModPhys.81.1827}{Rev.~Mod.~Phys.~\textbf{81},
1827 (2009)}.

\bibitem{PirozhenkoLambrecht} I.~Pirozhenko and A.~Lambrecht,
\emph{Influence of slab thickness on the Casimir force},
\href{https://doi.org/10.1103/PhysRevA.77.013811}{Phys.~Rev.~A \textbf{77},
013811 (2008)}.


\bibitem{PhilbinLeonhardt} T.~G.~Philbin and U.~Leonhardt,
\emph{No quantum friction between uniformly moving plates},
\href{https://doi.org/10.1088/1367-2630/11/3/033035}{New J.~Phys.~\textbf{11},
033035 (2009)}.

\bibitem{VolokitinComment} A.~I.~Volokitin and B.~N.~J.~Persson,
\emph{Comment on ``No quantum friction between uniformly moving plates''},
\href{https://doi.org/10.1088/1367-2630/13/6/068001}{New J.~Phys.~\textbf{13},
068001 (2011)}; T.~G.~Philbin and U.~Leonhardt, \emph{Reply},
\href{https://doi.org/10.1088/1367-2630/13/6/068002}{\textbf{13}, 068002
(2011)}.

\bibitem{KlattBuhmann} J.~Klatt, R.~Bennett, and S.~Y.~Buhmann,
\emph{Spectroscopic signatures of quantum friction},
\href{https://doi.org/10.1103/PhysRevA.94.063803}{Phys.~Rev.~A \textbf{94},
063803 (2016)}.


\bibitem{HartmannIngold} M.~Hartmann, G.-L.~Ingold, and P.~A.~Maia Neto,
\emph{Plasma versus Drude modeling of the Casimir force: Beyond the proximity
force approximation},
\href{https://doi.org/10.1103/PhysRevLett.119.043901}{Phys.~Rev.~Lett.~\textbf{119},
043901 (2017)}.




\end{thebibliography}
\end{document}